\documentclass[english,final]{siamltex}

\usepackage{amsmath,amssymb,epsf}
\usepackage{graphicx}
\usepackage{color}
\usepackage{latexsym}
\usepackage{changebar}
\usepackage{array}
\usepackage{subfigure}
\usepackage{babel}
\usepackage{cite}
 \newtheorem{thm}{Theorem}[section]
 
  \newtheorem{rem}{Remark}

\definecolor{grau}{gray}{.5}
\definecolor{schwarz}{gray}{0}

\DeclareGraphicsExtensions{.png,.eps,.ps,.jpg,.jpeg}

\newcommand{\reff}[1]{(\ref{#1})}
\newcommand{\ol}[1]{\overline{#1}}

\newcommand{\cg}[1]{\mathcal{#1}}

\newcommand{\av}[1]{\left|#1\right|}

\newcommand{\N}[2]{\left\|#1\right\|_{#2}}
\newcommand{\brkts}[1]{\left(#1\right)}
\newcommand{\ebrkts}[1]{\left[#1\right]}

\newcommand{\brcs}[1]{\left\{#1\right\}}

\newcommand{\abrkts}[1]{\langle #1\rangle}

\newcommand{\pd}[2]{\frac{\partial #1}{\partial #2}}

\newcommand{\bsplitl}[2]{
\begin{equation}
\begin{split}
#1
\end{split}
\label{#2}
\end{equation}}
\newcommand{\bsplit}[1]{
\begin{equation*}
\begin{split}
#1
\end{split}
\end{equation*}}

\begin{document}

\title{Homogenization of the Poisson-Nernst-Planck equations for ion transport in charged porous media
\thanks{This work was supported by the Swiss National Science Foundation (SNSF) under 
the grant PBSKP2-12459/1 (MS) and in part by the National Science Foundation under contract DMS-0948071 (MZB).}}

\date{\today{}}

\author{Markus Schmuck\thanks{
	Maxwell Institute for Mathematical Sciences and Department of Mathematics, Heriot-Watt University, Edinburgh, EH14 4AS, UK
	}
        \and 
        Martin Z. Bazant \thanks{Departments of Chemical Engineering and Mathematics, Massachusetts Institute of Technology, Cambridge, MA 02139, USA}
        }


\maketitle

\begin{abstract}
Effective Poisson-Nernst-Planck (PNP) equations are derived for macroscopic ion transport in charged porous media under 
periodic fluid flow by an asymptotic multi-scale expansion with drift. The microscopic setting is a two-component periodic composite consisting of a dilute electrolyte continuum (described by standard PNP equations) and a continuous dielectric matrix, which is impermeable to the ions and carries a given surface charge.   Four new features arise in the upscaled equations: (i) the effective ionic diffusivities and mobilities become tensors, related to the microstructure; (ii) the effective permittivity is also a  tensor, depending on the electrolyte/matrix permittivity ratio and the ratio of the Debye screening length 
to the macroscopic length of the porous medium; (iii) the microscopic fluidic convection is replaced by 
a diffusion-dispersion correction in the effective diffusion tensor; and (iv)  the surface charge per volume appears as a continuous ``background charge density", as in classical membrane models.  The coefficient tensors in the upscaled PNP equations can be calculated from periodic reference cell problems. For an insulating solid matrix, all gradients are corrected by the same tensor, and the Einstein relation holds at the macroscopic scale, which is not generally the case for a polarizable matrix, unless the permittivity and electric field are suitably defined.  In the limit of thin double layers, Poisson's equation is replaced by macroscopic electroneutrality (balancing ionic and surface charges). The general form of the macroscopic PNP equations may also hold for concentrated solution theories, based on the local-density and mean-field approximations. These results have broad applicability to ion transport in porous electrodes, separators, membranes, ion-exchange resins, soils, porous rocks, and biological tissues.
\end{abstract}

\begin{keywords} 
diffusion, electromigration, porous media, membranes, Poisson-Nernst-Planck equations, homogenization
\end{keywords}

\begin{AMS}
\end{AMS}

\pagestyle{myheadings}
\thispagestyle{plain}
\markboth{M. Schmuck and M. Z. Bazant}{Ion transport equations in porous media}

\section{Introduction}\label{sec:Intro}

The theory of electrochemical transport in free solutions is well 
developed~\cite{rubinstein_book,probstein1989,Newman2004}, but in many practical situations, 
ions move through porous microstructures with internal surface charge.  Important examples in 
biology include nerve impulse propagation in the porous intracellular matrix of an axon~\cite{weiss_book}, 
selective ion transport through protein-based ion channels in cell membranes~\cite{roux2004,eisenberg1998}, 
and the electroporation of porous tissues for drug delivery and medical diagnostics~\cite{weaver1996}.  
In chemical engineering, the selective transport of ions and charged particles through membranes, gels and 
porous media is widely used for particle separations~\cite{giddings91},  desalination and ion exchange ~\cite{probstein1989,helfferich1962}, characterization of porous rocks~\cite{sahimi1995}, energy conversion in fuel 
cells~\cite{ohare2009} and energy storage in batteries~\cite{newman1975} and electrochemical 
supercapacitors~\cite{conway1999}. Analogous nanoscale transport phenomena are also beginning 
to be exploited in microfluidic devices~\cite{schoch2008,bruus2008}, which involve artificial 
porous structures with precisely controlled geometries and surface properties.

In {\it microscopic} continuum models of electrolytes, the ionic fluxes are given by the Nernst-Planck equations describing diffusion and electromigration in the mean electric field, which is determined self-consistently from the mean ionic charge density via Poisson's equation. The resulting Poisson-Nernst-Planck (PNP) system has been studied extensively for the past century in the dilute solution approximation, not only for electrolytes \cite{eisenberg1999,bazant2004,bazant2005}, but also for semiconductors, where electrons and holes behave like anions and cations, respectively~\cite{markowich2002}. The dilute-solution PNP equations can be derived rigorously from stochastic Langevin equations for the motion of point-like ions~\cite{nadler2004}. 

Recently, a variety of modified PNP equations for concentrated solutions have been developed to describe strong interactions of finite-sized ions with charged surfaces at the nanoscale, as reviewed by ~\cite{Bazant2009}. Hard-sphere density functional theory ~\cite{gillespie2002,gillespie2011} and simpler mean-field models~\cite{Kilic2007a,olesen2010} have been used  to modify the Nernst-Planck equations for ionic fluxes to account for steric hindrance.  Poisson's equation has also been modified to account for electrostatic correlations ~\cite{santangelo2006,weeks2006,hatlo2010,bazant2011,storey2012}, explicit treatment of solvent dipoles~\cite{koehl2009} and solvation energy variations due to nonuniform permittivity~\cite{wang2010}.  All of these developments improve the microscopic description of ion transport close to charged surfaces, but our focus here is on the homogenization of such models over a charged microstructure to derive effective PNP equations valid at the {\it macroscopic} scale.  

There is a long history of heuristic models for macroscopic ion transport in charged membranes and porous media, dating back at least to the 1930s~\cite{teorell1953}. A classical concept in membrane science, which we place on a rigorous footing below for general porous media, is the notion of a fixed ``background charge" entering Poisson's equation, due to the volume-averaged surface charge of the porous medium~\cite{helfferich1962}. In nanoporous membranes, the double layers are thick compared to the pore thickness, so that there are only small variations in diffuse ionic charge between the fixed surface or molecular charges. For most porous media, however, the double layers are assumed to be thin, leaving the pore spaces to be mostly filled with neutral solution, and Poisson's equation is replaced by electroneutrality, without accounting for the background charge. In electrochemistry,  this is a fundamental assumption of ``porous electrode theory" (PET), introduced by Newman and Tobias~\cite{newman1962}, which postulates electroneutrality within the pores and effective Nernst-Planck equations of the same form as in the bulk solution, except for an empiricial tortuosity factor multiplying the ionic diffusivities. This approach has been applied extensively to batteries~\cite{newman1975,Newman2004,dargaville2010,lai2011,ferguson2012}. The nonlinear effects of double layer charging~\cite{bazant2004} have also recently been incorporated into PET to model capacitive desalination and energy storage~\cite{biesheuvel2010,biesheuvel2011}.  The assumptions of PET have been tested against large-scale numerical solutions of the microscopic transport equations in certain cases of realistic microstructures~\cite{garcia2005,garcia2007}, but mathematical derivations are still be needed to predict the form of the macroscopic equations and to provide a systematic method to calculate their coefficients. This is the goal of  the present work. 

To the best of our knowledge, this seems to be the first attempt of systematically 
upscaling fully nonlinear and time dependent PNP equations in charged porous microstructures. 
In spite of the many important applications listed above, there has only been 
recently increasing interest in the systematic upscaling of PNP equations. Up to now, most derivations \cite{Jackson1963,Gross1968,Holmes1990} have been formal in nature and require simplifying assumptions, such as a neutral bulk 
and a linearized PNP problem or even the equilibrium Poisson-Boltzmann approximation, for instance. Moyne and Murad~\cite{moyne2006} assume a Boltzmann equilibrium distribution of ions in a binary electrolyte at the pore scale and perform a homogenization analysis to derive effective equations for deformable porous media.  For neutral species, the homogenization of linear diffusion over porous microstructures is well developed, and rigorous bounds are available for the effective macroscopic diffusivity tensor over all possible microstructures \cite{torquato2002}. Looker and Carnie~\cite{looker2006} make the same approximation of microscopic Boltzmann equilibrium and derive symmetric Onsager relations for linear response, without stating the general effective equations at the macroscopic scale. Allaire et al.~\cite{allaire2010} revisit the derivation of Looker and Carnie~\cite{looker2006} using two-scale convergence methods developed by Nguetseng~\cite{nguetseng1989} and Allaire~\cite{allaire1992} and 
prove the positive definiteness of the Onsager tensor, which requires the explicit use of an electroneutrality assumption and linearized equations. 
First bounds on the error arising between the full, nonlinear, microscopic, periodic porous media problem and the upscaled/homogenized approximation can be found in \cite{Schmuck2012}. Very recently, one can also find upscaling results for the full Navier-Stokes-PNP system \cite{Ray2012,Schmuck2010}. For rigorous analytical results on the full Stokes-Nernst-Planck-Poisson system we refer to \cite{Jerome2002,Schmuck2009} and for according reliable, efficient and convergent numerical schemes and computational methods to \cite{Prohl2010}.

In this article, we derive porous-media PNP equations for charged microstructures using the method of multiple scale expansion with drift. In contrast to Refs. \cite{Schmuck2010,Schmuck2012}, we account for 
crucial, nonlinear influence of surface charge on the pore walls in the PNP equations, and for
periodic fluid flow defined on a periodic reference cell,  thus going well beyond the analysis of Ref. \cite{Schmuck2010}. 
The resulting macroscopic transport equations have the following general form:
\bsplitl{
\textrm{\bf Homogenized PNP System:}\,\,\,
\begin{cases}
\theta \partial_t   c_0^\pm
	 = {\rm div}\brkts{
		\hat{\rm D}({\bf v})\nabla c_0
+ z_\pm c_0^\pm \hat{\rm M} \nabla \phi_0  }
\\
-{\rm div}\brkts{
		\hat{\epsilon}^0\phi_0
	}
	=\brkts{c_0^+ - c_0^-}
	+\rho_s
\end{cases}
}{upscSys}
where $c_0^+$ and $c_0^-$ are the densities of positively and negatively charged ions, 
respectively, $z_\pm$ are the charge numbers (ion valences with sign), $\phi_0$ is the electrostatic potential, $\theta$ is the porosity, 
and the effective porous media correction tensors $\hat{\rm D}({\bf v}),\,\hat{\rm M}$ and $\hat{\epsilon}^0$ for the diffusivity, mobility, and permittivity, respectively, are defined in \reff{EfTe}. 
The case without fluid flow is obtained by setting ${\bf v}={\bf 0}$. 
In the limit of thin double layers for isotropic media, our equations are physically equivalent to those proposed in recent work\cite{Ramirez2003,He2009,Szymczyk2010,Mani2009,dydek2011,dydek2013,yaroshchuk2012,deng2013} based on intuitive and physical reasoning for nanochannels or porous media, where the potential is determined implicitly by macroscopic electroneutrality, including not only the ions, but also the surface charge. 
Here, we derive more general PNP equations, valid for any double layer thickness, which preserve the form of Poisson's equation with a modified effective permittivity, where the electric field is produced by the total charge density. Our multiscale 
approach allows us to systematically calculate the tensorial coefficients in the macroscopic equations accounting for 
different pore geometries defined by a periodic reference cell.
By including locally periodic fluid flow, we also
obtain a set of so-called diffusion-dispersion relations, which generalize in some ways the classical approximation of Taylor-Aris dispersion~\cite{yaroshchuk2011}. The key assumption of local diffusive quasi-equilibrium holds in many situations, but not at very high currents (exceeding macroscopic diffusion limitation) where  fast electro-osmotic surface convection leads to incomplete local mixing at the pore scale~\cite{dydek2011,rubinstein2013,deng2013}.

\medskip
 
 The article is organized as follows. We begin in Section~\ref{sec:HoMe} by recalling the PNP equations for homogeneous bulk solutions. In Section~\ref{sec:MiFo} we extend this coupled system towards a microscopic formulation in porous media . 
We state our main result of effective macroscopic (Stokes-)PNP equations in Section~\ref{sec:MaRe}. A formal proof by 
 the multiple scale method with drift follows in Section~\ref{sec:Pr}. We investigate physical implications of the new effective macroscopic Stokes-PNP system in Section~\ref{sec:Imp}. That means,                
we  briefly discuss the  effective diffusivity and mobility tensors and investigate the validity of Einstein's relation between them. We state conditions under which our results allow for an analytical computation of effective porous media coefficients in 
the case of straight channels in Section \ref{sec:StCh} and briefly exemplify irregular channels in the same Section.  We discuss definitions of tortuosity in Section \ref{sec:Tor}  and derive the general ambipolar diffusion equation for  a binary electrolyte in a charged porous medium in Section \ref{sec:AmDi}. In Section \ref{sec:TDL}, we take the limit of  thin double layers in the porous-media PNP equations.  In Section \ref{sec:optCond}, we suggest an approximate microstructural optimization of the effective conductivity of a symmetric binary electrolyte for parallel straight channels. In Section \ref{sec:Disc}, we  conclude by discussing possible extensions and applications of our homogenized PNP equations.

\subsection{Homogeneous media: Basic theory}\label{sec:HoMe}
 We adopt the well studied mathematical framework for dilute binary electrolytes  \cite{bazant2004,Bazant2009,Chu2007,Kilic2007,Kilic2007a,olesen2010}. For simplicity, we restrict ourselves to the symmetric 
 case $z=z^+=z^-$, $D=D_+=D_-$, and $M=M_+=M_-$ during the upscaling. An extension towards dilute, asymmetric binary electrolytes with arbitrary ionic charges $q_{\pm}=\pm z_\pm e$, diffusivities $D_{\pm}$, and mobilities $M_{\pm}$ is subsequently considered in Section \ref{sec:AmDi}.  The variables $z,\,z^+,\,$ and $z^-$ refer to valences of ions and $e$ denotes the elementary charge. We motivate that generalizations towards incompressible fluid flow are studied analytically and computationally in \cite{Jerome2002,Schmuck2009,Prohl2009,Prohl2010}. All equations subsequently considered are defined in a bounded, convex, and connected domain $\Omega\subset\mathbb{R}^N$ with $1\leq N\leq 3$.

The  concentrations of positively and negatively charged ions 
$c^\pm(x,t)$ evolve according to mass conservation laws 
\bsplitl{
\partial_tc^\pm
	= - {\rm div}\brkts{- c^\pm M_\pm\nabla\mu_\pm}\,,
}{NP}
where the classical Nernst-Planck fluxes (in parentheses) are expressed according to linear irreversible thermodynamics in terms of the gradients of the diffusional chemical potentials
$\mu_\pm$, given by the
\bsplitl{
\textrm{\bf Dilute Solution Theory:}\qquad
\mu_\pm 
	= kT{\rm ln}\,c^\pm
	+z_\pm e\phi\,.
}{ChPo}
The variable $\phi$ is the electrostatic potential, which describes 
the Coulomb interaction in a mean-field approximation. $k$ denotes the Boltzmann constant, $T$ the absolute temperature, 
and $e$ the elementary charge. The 
coefficients $D_\pm$ are the (tracer) diffusivities of the two 
ionic species. The mobilities, $M_\pm$, which give the drift velocity in response to an applied force, are then obtained by Einstein's relation $M_\pm = \frac{D_\pm}{kT}$, which must hold for individual ions by the fluctuation-dissipation theorem. The total mean ionic 
charge density $\rho$ controls the spatial variation of the 
potential $\phi$ through Poisson's equation,
\bsplitl{
-\epsilon_s\Delta\phi 
	= \rho
	:= ze(c^+-c^-)\,,
}{P}
where $\epsilon_s$ is the dielectric permittivity of the solution (roughly equal to that of the solvent),  
assumed to be a constant.

Next, we cast the equations in a dimensionless form 
using $\ell$ as a reference length scale and 
$t_D=\ell^2/D$ as the reference time scale. We use the 
thermal voltage $\frac{kT}{e}$ as a scale for the electric 
potential. We introduce the reduced variables 
\bsplitl{
\tilde{c}^+=\frac{c^+}{\ol{c}}\,,
\qquad
\tilde{c}^-=\frac{c^+}{\ol{c}}\,,
\qquad
\tilde{\phi} = \frac{ze\phi}{kT}
\qquad
\tilde{x} = \frac{x}{\ell}\,,
\qquad
\tilde{t} = \frac{t}{t_D}\,,
\qquad
\tilde{\nabla} = \ell\nabla\,,
}{RV}
where $\ol{c}$ is a reference concentration of ions, such as the nominal salt concentration of a quasi-neutral bulk electrolyte 
obtained from a large enough reservoir next to $\Omega$, i.e., prior to its perfusion in the porous medium $\Omega$. The reference solution could be removed, or maintained in contact with the porous medium. We thus arrive at dimensionless Poisson-Nernst-Planck equations containing only  the 
dimensionless parameter $\ol{\epsilon}=\frac{\lambda_D}{\ell}$, 
\bsplitl{
\partial_{\tilde{\ol t}}\tilde{c}^+
	& = \tilde{\rm{div}}\brkts{
		\tilde{\nabla}\tilde{c}^+
		+\tilde{c}^+\tilde{\nabla}\tilde{\phi}
	}\,,
\\
\partial_{\tilde{t}}\tilde{c}^-
	& = \tilde{\rm{div}}\brkts{
		\tilde{\nabla}\tilde{c}^-
		-\tilde{c}^-\tilde{\nabla}\tilde{\phi}
	}\,,
\\
-\ol{\epsilon}^2\tilde{\Delta}\tilde{\phi}
	& = \tilde{c}^+-\tilde{c}^-\,,
}{PNP}
where $\ol{\epsilon}:=\frac{\lambda_D}{\ell}$ is a dimensionless parameter defined by the Debye screening length $\lambda_D:=\brkts{\frac{\epsilon_skT}{2e^2\ol{c}}}^{1/2}$ for a symmetric binary electrolyte. 


In our analysis below, we shall use dimensionless equations and drop the tilde accents for ease of notation.

\subsection{Porous media: Microscopic formulation}\label{sec:MiFo}

\begin{figure}
	\centering
	\includegraphics[height=6.5cm,width=13cm]{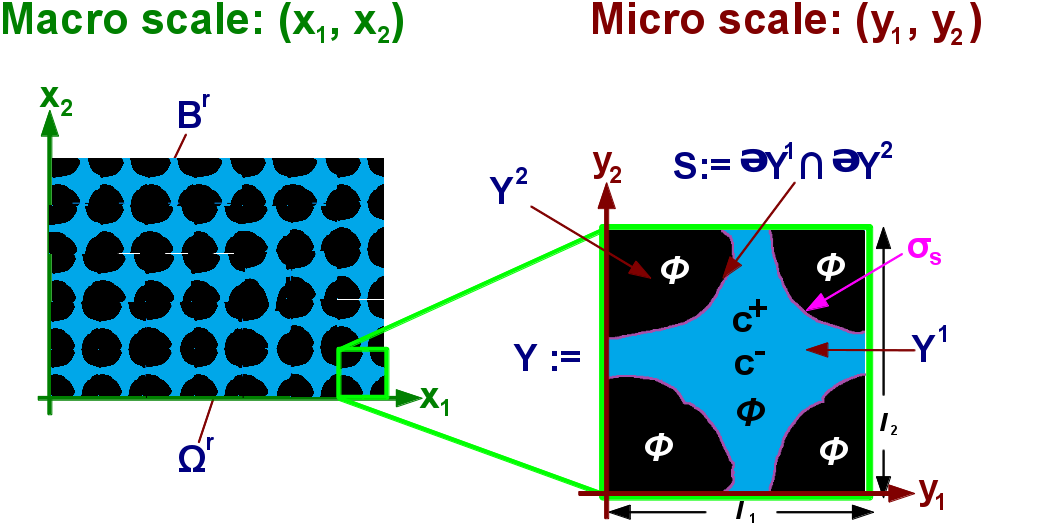}
	\caption{{\bf Left, macro scale:} Domain $\Omega:=\Omega^r\cup B^r$ with the solid-liquid interface 
	$I^r:=\partial\Omega^r\cap\partial B^r$. {\bf Right, micro scale:} Periodic reference cell 
	$Y:=Y^1\cup Y^2:=[0,l_1]\times[0,l_2]$ with solid-liquid interface $S:=\partial Y^1\cap\partial Y^2$. In 
	case $\sigma_{\rm s}\neq 0$, then $S$ is assumed to be smooth.
	The scaling parameter $r$ defines the microscale by $y=x/r$ and measures the characteristic size of the heterogeneities. The upscaling then consists in passing to the limit $r\to 0$, i.e., the electrolyte and the solid phase are homogeneously 
	mixed while keeping the corresponding volume fractions constant.}
	\label{fig:MicMac}
\end{figure}

Here, we extend the system \reff{PNP} towards fluid flow and perforated domains $\Omega^r\subset\mathbb{R}^N$
instead of the homogeneous $\Omega\subset\mathbb{R}^N$ from Section \ref{sec:HoMe}. The dimensionless parameter $r>0$ is
defined by $r=\frac{l}{L}$ where $l$ 
represents
the characteristic pore size and $L$ is the characteristic length of the porous medium, see Figure \ref{fig:MicMac}. The
pores are defined by a single, periodic reference cell $Y:= [0,l_1]\times[0,l_2]\times\dots\times[0,l_N]$, $l_i\in\mathbb{R}^N_{\geq 0}$, which defines the characteristic pore geometry of the porous medium. Herewith, the characteristic pore size can be defined by $l:=\sqrt{l_1^2+l_2^2+\dots+l_N^2}$ for instance. A well-accepted approximation is to periodically cover the
macroscopic porous medium by the characteristic reference cell $Y$, see Figure \ref{fig:MicMac}. The pore
and the solid phase
of the medium are denoted by $\Omega^r$ and $B^r$, respectively.
These sets are defined by,
\bsplitl{
\Omega^r
    & := \bigcup_{{\bf z}\in\mathbb{Z}^N}r\brkts{Y^1+{\bf z}}\cap\Omega\,,
\qquad
B^r
    := \bigcup_{{\bf z}\in\mathbb{Z}^N}r\brkts{Y^2+{\bf z}}\cap\Omega
    =\Omega\setminus\Omega^r\,,
}{Oe2}
where the subsets $Y^1,\,Y^2\subset Y$ are defined such that $\Omega^r$
is a connected set. The domain $\Omega\subset\mathbb{R}^N$ in \reff{Oe2} is an open, bounded, connected and convex subset for $1\leq N\leq 3$. We denote by $I^r:=\Omega^r\cap B^r$ the solid-liquid interface 
that is assumed to be smooth if there is no surface charge $\sigma_{\rm s}$ present, i.e., $\sigma_{\rm s}\neq 0$. Correspondingly, $Y^1$ stands for the pore phase (e.g. liquid or gas phase),
see Figure \ref{fig:MicMac}.

We use dimensionless length and time variables (\ref{RV}) scaled to the length and time scales for diffusion across a homogeneous medium of characteristic length $\ell$ which allows us to vary the Debye length independent of the upscaling/homogenization and relative to a suitably chosen length scale such as the microscopic pore size $l$ or the macroscopic length $L$ of the porous medium satisfying $l\leq\ell\leq L$ and $l\ll L$. This ensures that the porous media approximation (by homogenization) is not affected by an additional thin or thick double layer approximation.

Hence, in deriving effective macroscopic PNP equations we work with  the following macroscopic dimensionless variables:
\begin{equation}
\epsilon = \frac{\lambda_D}{\ell}\,, 
\ \  \ 
x = \frac{\ol x}{\ell}\,, 
\ \ \ 
\nabla = \ell\ol\nabla\,, 
\ \  \ 
t = \frac{\ol t D}{\ell^2} \,.
\label{macro}
\end{equation}
Since we consider here the case of charged porous media, the reference bulk salt concentration $\overline{c}$ will be replaced with the 
averaged surface charge $$\rho_s:=\frac{1}{\av{Y}}\int_{\partial Y^1\cap\partial Y^2}\sigma_{\rm s}(y)\,d\sigma(y),$$
where $\sigma_{\rm s}(y)$ is Y-periodic.
As in the case without fluid flow \cite{Schmuck2013}, we assume scale separated chemical potentials in the sense of

\medskip

{\bf Assumption (AI):} (Scale separation) \emph{We assume that the chemical potentials $\mu_+(c_0^+,\phi_0)$ and $\mu_-(c_0^-,\phi_0)$ are scale separated, that means,
\bsplit{
\frac{\partial \mu_\iota(c_0^\iota,\phi_0)}{\partial x_k}
	= \begin{cases}
\quad
		0 & \text{on the reference cell $Y$}\,,
\\\quad
		\frac{\partial \mu_\iota(c_0^\iota,\phi_0)}{\partial x_k} & \text{on the macroscale $\Omega$}\,,
	\end{cases}
}
for $\iota=+,-$, where $c_0^\pm({\bf x})$ and $\phi_0$ are the upscaled/slow variables solving the upscaled PNP 
system.
}
%
%
%
%

\medskip

We additionally consider periodic fluid flow which is assumed 
to induce a dominant convection by the 

\medskip

{\bf Assumption (AII):} \emph{Suppose that the P\'eclet number satisfies}
\bsplitl{
{\rm Pe} \sim \frac{{\rm Pe_{loc}}}{r}\,.
}{PeSc}

\medskip

Characterization {\bf (CII)} describes the situation of dominant convection, i.e., $V\sim 1/r$. Before we can formulate our next assumption, we need to introduce the 
function spaces 
\bsplitl{
V^1(\Omega)
	:=\brcs{
		u\in H^1(\Omega)\,\bigr|\,\int_\Omega u\,d{\bf x}=0
	}\,,
\\
H^1_\sharp(Y)
	:=\ol{\brcs{
		u\in C^\infty(Y)\,\biggr|\,u(y) \textrm{ is }Y-\textrm{periodic and }\frac{1}{|Y|}\int_Y u(y)\,dy=0
	}}^{H^1}\,,
\\
H^1_\sharp({\rm div},Y)
	:=\ol{\brcs{
		{\bf u}\in H^1_\sharp(Y,\mathbb{R}^N)\,\biggr|\,{\rm div}\,{\bf u} = 0
	}}^{H^1}\,,
}{W12per}
where $\ol{\brcs{\cdot}}^{H^1}$ denotes the closure with respect to the $H^1$-norm. In a corresponding way, we 
apply the notation $L^\infty_\sharp(Y)$ below.

\medskip

{\bf Assumption (AIII):} \emph{The convective velocity}
\bsplitl{
{\bf u}_r (x,t)
	=
	{\bf v}(x/r)\,,
}{PeVe}
\emph{is periodic with a divergence-free vector field ${\bf v}(y)\in H^1_{\sharp}({\rm div},Y^1)$ satisfying} 
\bsplitl{
\max_{y\in\ol{Y}^1} \av{{\bf v}(y)}
	\leq C\,,
	\qquad
	{\rm div}_y\,{\bf v}(y)
	=0
	\quad\textrm{in }Y^1\,,
\\
{\bf v}(y)\cdot\pmb{\nu}(y)
	=0
	\quad\textrm{on }
	\partial Y^2\,.
}{v}
\medskip

We also assume that the interface is smooth enough in order to 
allow for a surface two-scale limit \cite{Allaire1996}. Hence, we make the following 

\medskip
{\bf Assumption (AIV):} \emph{If the surface charge density $\sigma_{\rm s}$ is non-zero, 
the we suppose that the interface $S=I^r\cap rY$ is smooth. Moreover, the initial conditions for the concentrations 
satisfy the compatibility condition (global electroneutrality):
\bsplitl{
\int_{\Omega^r}{\rm c}^+-{\rm c}^-\,dx
	= \int_{I^r}\sigma_{\rm s}\,do(x)\,.
}{SuChCo}
}
\medskip

We collect the macroscopic boundary conditions in the 

\medskip
{\bf Assumption (AV):} (Academic boundary conditions) \emph{Let the field vector ${\bf u}_r :=[{\bf v},{\rm c}_r^+,{\rm c}_r^-,\phi_r]'$, which solves \reff{MiPr}, satisfy homogeneous Dirichlet boundary conditions on $\partial\Omega$ for the quantities ${\rm u}_r^\iota$ 
with $\iota=1,2,3$ and no-flux boundary conditions for ${\rm u}_r^4$.
}
\medskip

Finally, the Assumption {\bf (AII)}--{\bf (AIV)} allow us to formulate the microscopic porous media formulation as 
follows 
\bsplitl{
\textrm{\bf (micro) }
\begin{cases}
-\Delta {\bf v} +\nabla_y q 
	=
	- {\bf e}_1 
	&\qquad\textrm{in }Y^1\,,
\\
\qquad
{\rm div}_y\, {\bf v} = 0
	&\qquad\textrm{in }Y^1\,,
\\
\qquad
{\bf v} = 0
	&\qquad\textrm{on }\partial Y^1\,,
\\
\qquad
{\bf v},\, q
	&\qquad\textrm{are $Y$-periodic\,,}
\\
\frac{\partial c_r^\pm}{\partial t}
	= {\rm div}\brkts{
		\nabla c_r^\pm
		\pm c_r^\pm\nabla\phi_r
		-\frac{{\rm Pe}_{loc}}{r}\,{\bf v}(x/r,t)c^\pm_r
	}
	&\qquad\textrm{in }\Omega^r_T\,,
\\
\qquad
{\bf j}^\pm_r\cdot\pmb{\nu}
	= 0
	&\qquad\textrm{on }\partial\Omega^r_T\setminus\partial\Omega\,,
\\
-{\rm div}\brkts{
		\varepsilon(x/r)\nabla\phi_r
	}
	= {\rm c}^+_r-{\rm c}^-_r
	&\qquad\textrm{in }\Omega_T\,,
\\
\qquad
\phi_r\bigr|_{\Omega^r}
	=  \phi_r\bigr|_{B^r}
	&\qquad\textrm{on }I_r\times]0,T[\,,
\\
\qquad
(\pmb{\nu}\cdot\nabla)\phi_r
	= r\rho_{\rm s}(x/r)
	&\qquad\textrm{on }I_r\times]0,T[\,,
\end{cases}
}{MiPr}
where ${\bf u}_r:=\frac{\tilde{\bf u}_r}{V}$ is the dimensionless fluid velocity for a reference velocity $V:=\av{\bf v}$. 
We assume a horizontal flow induced by ${\bf e}_1$ as in \cite{Allaire2010a} for pure convection diffusion problems.
$\pmb{\nu}$ denotes the normal on $I^r$ pointing outward of the 
pore phase $\Omega^r$ and $\sigma_{\rm s}(x/r)$ is $Y$-periodic. We denote by ${\rm c}_r^\pm$ the trivial extension by zero of the concentrations 
$c^\pm$ in the Poisson equation \reff{MiPr}$_7$.
The parameter ${\rm Pe}:=\frac{L V}{D}$ is the dimensionless P\'eclet number and 
${\bf j}^\pm_r:=\nabla c_r^\pm
		\pm c_r^\pm\nabla\phi_r
		-{\rm Pe}_{loc}/r\,{\bf u}_rc^\pm_r$ represents the flux of positive and negatively charged species.
Finally, we recall from \cite{Schmuck2010,Schmuck2012} that $\varepsilon(x):=\epsilon^2\chi_{\Omega^r}(x)+\alpha\chi_{\Omega\setminus\Omega^r}(x)$ with the dimensionless 
dielectric permittivity $\alpha:=\frac{\epsilon_m}{\epsilon_f}$ where 
\bsplitl{
\chi_{\omega}(x)
	= \begin{cases}
	1 & x\in\omega\,
	\\
	0 & \textrm{else}\,,
	\end{cases}
}{chi}
for $\omega\in\brcs{\Omega^r,B^r}$
and $\chi_\omega(x/r)=\chi_\omega(y)$ is $Y$-periodic.
The variables $\epsilon_m$, and 
$\epsilon_f$ are the dielectric constants of the porous medium and of the electrolyte, respectively. For simplicity, we consider 
no-flux boundary conditions with respect to ion densities.

\medskip
\begin{rem}\label{rem:ExtConc}
We note that in \reff{MiPr}$_7$, the concentrations $c^\pm$ are extended in the solid phase $B^r$ by 
an abstract extension operator $T_s$ as introduced in \cite{Acerbi1992} for instance.
\end{rem}
\medskip

In the next section, we present our main result.

\section{Main results: Effective macroscopic porous media approximation}\label{sec:MaRe}
\begin{figure}[htbp]
\begin{center}
\setlength{\unitlength}{0.8cm}
\begin{picture}(14,6)(0,0)					
\thicklines
\multiput(0.0,0.0)(0.8,0.0){6}{\framebox(0.8,0.8)[s]{\put(0.4,0.0){\circle*{0.6}}}}
\multiput(0.0,0.8)(0.8,0.0){6}{\framebox(0.8,0.8)[s]{\put(0.4,0){\circle*{0.6}}}}
\put(-0.5,1.6){\line(1,0){0.5}}
\multiput(0.0,1.6)(0.8,0.0){6}{\framebox(0.8,0.8)[s]{\put(0.4,0){\circle*{0.6}}}}
\put(-0.5,2.4){\line(1,0){0.5}}
\put(-0.25,1.6){\vector(0,1){0.8}}
\put(-0.25,2.4){\vector(0,-1){0.8}}
\put(-0.6,1.9){$r$}
\multiput(0.0,2.4)(0.8,0.0){6}{\framebox(0.8,0.8)[s]{\put(0.4,0){\circle*{0.6}}}}
\multiput(0.0,3.2)(0.8,0.0){6}{\framebox(0.8,0.8)[s]{\put(0.4,0){\circle*{0.6}}}}
\put(0,5.0){Strongly heterogeneous}
\put(0,4.5){material, i.e. $0<r\ll 1$}
\put(6.4,2.25){($r\to 0$)}
\put(6.0,2.0){\vector(2,0){2.0}}
\put(9.5,0.0){\fcolorbox{black}{grau}{\makebox(4.8,4.0)[]{}}} %
\put(9.3, 4.5){Homogenous approximation}
\end{picture}
\caption{{\bf Left:} A composite material whose characteristic heterogeneity has the length $r$. 
{\bf Middle:} Passing to the limit $r\to 0$ under constant volume fraction between circle and square. 
{\bf Right:} The limit problem is obtained by the method of an asymptotic multi-scale expansion with drift.} 
\label{fig:3.1}
\end{center}
\end{figure}
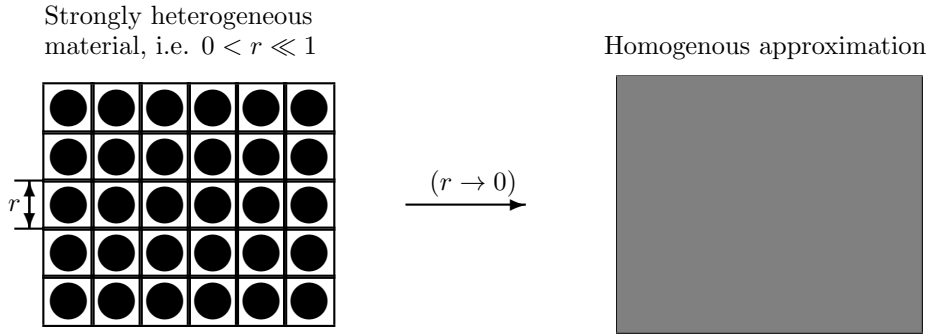

Based on the microscopic considerations in the previous Section 
\ref{sec:MiFo}, we can immediately state our main result which is the upscaling 
of \reff{MiPr} by a multiscale expansion with drift as depicted in Figure \ref{fig:3.1}.

\medskip
\begin{thm}\label{thm:pmStPNP} Under the Assumptions {\rm (AI)}--{\rm (AV)} and for 
$\sigma_{\rm s}\in L^\infty_\sharp(Y)$, the solution ${\bf u}_r:=[{\bf v},{\rm c}_r^+,{\rm c}_r^-,{\phi}_r]'$ 
of problem \reff{MiPr} admits for $\iota=2,3,4$ the leading order representation 
\bsplitl{
{\rm u}_r^\iota
	= {\rm u}_0^\iota
	-r\sum_{j=1}^N\xi^{\iota_j}(x/r)\frac{\partial {\rm u}_0^\iota}{\partial x_j}
	+{\cal O}(r^2)
	\qquad\textrm{for }\iota=2,3,4\,,
}{AsExTh}
where $\xi^{\iota_j}(y)\in H_\sharp^{1}(Y^1)$ for $\iota=2,3$ and $\xi^{3_j}(y)\in H_\sharp^{1}(Y)$ 
solve the following reference cell problems in the distributional sense,
\bsplitl{
\begin{cases}
-\sum_{i,j=1}^N\pd{}{y_i}\brkts{
		\delta_{ij}\pd{}{y_j} 
			\xi^{\iota_j}(y) 
	}
	+{\rm Pe}_{loc}({\bf v}\cdot\nabla_y)\xi^{\iota_j}
\\\qquad
	=-\sum_{i,j=1}^N\pd{}{y_i}\brkts{
		\delta_{ij}\pd{}{y_j} 
			\xi^{3_j}(y) 
	}
	+({\rm v}^*_j-{\rm Pe}_{loc}{\rm v}_j)
	&\textrm{in }Y^1\,,
\\\quad
	\pmb{\nu}\cdot\brkts{
		\brkts{
			\nabla_y\xi^{\iota_j}(y)-{\bf e}_i
		}
		+\brkts{
			\nabla_y\xi^{3_j}(y)-{\bf e}_i
		}
		+{\rm Pe}_{loc}{\bf v}\xi^{\iota_j}
	}=0
	&
	\textrm{on }\partial Y^1\,,	
\\
-\sum_{i,j=1}^N\pd{}{y_i}\brkts{
		\varepsilon(y)\delta_{ij}\pd{}{y_j} 
			\xi^{3_j}(y) 
	}
	= 0
	&\textrm{in } Y\,,
\\
\end{cases}
}{ReCePr}
where $\delta_{ij}$ stands for the Kronecker $\delta$.
The field ${\bf u}_0:=[{\bf v},{\rm c}_0^+,{\rm c}_0^-,{\phi}_0]'\in H^1_\sharp({\rm div},Y^1)\times[H^1_0(\Omega)]^2\times V^1(\Omega)$ forms a solution of the following upscaled system,
\bsplitl{
\textrm{\bf Upscaled System:}\,\,\,
\begin{cases}
\theta\partial_t{\rm u}_0^\iota
	-{\rm div}\brkts{
		\hat{\rm D}({\bf v})\nabla {\rm u}^\iota_0
	}
	-{\rm div}\brkts{
		z_\iota{\rm u}_0^\iota\hat{\rm M}\nabla {\rm u}^3_0
	}
	=0
	&\textrm{in }\Omega_T\,,
\\
-{\rm div}\brkts{
		\hat{\epsilon}^0\nabla{\rm u}^3_0
	}
	=\brkts{{\rm u}^1_0-{\rm u}_0^2}
	+\rho_s
	&\textrm{in }\Omega_T\,,
\end{cases}
}{EffpmPNP}
where $\theta:=\frac{\av{Y^1}}{\av{Y}}$ defines the porosity, ${\bf v}$ solves \reff{MiPr}$_1$--\reff{MiPr}$_4$, and ${\rm u}_0^\iota$ satisfies the same boundary conditions as assumed in Assumption {\rm (AV)}. The diffusion-correction tensor 
$\hat{\rm D}({\bf v}):=\brcs{{\rm d}_{kl}({\rm v}_i)}_{1\leq k,l\leq N}$, the electro-convection tensor 
$\hat{\rm M}:=\brcs{{\rm m}_{kl}}_{1\leq k,l\leq N}$, and the effective permeability 
tensor $\hat{\epsilon}^0:=\brcs{\epsilon^0_{kl}}_{1\leq k,l\leq N}$ are defined by,
\bsplitl{
{\rm d}_{ik}({\rm v}_i)
	& := \frac{1}{\av{Y}}\int_{Y^1}\brkts{
		\delta_{ik}\brkts{1+{\rm Pe}_{loc}(\ol{\rm v}_i-{\rm v}_i)\xi^{\iota_k}}
		-\sum_{j=1}^N\delta_{ij}\frac{\partial\xi^{\iota_k}}{\partial y_j}
	}\,dy
	\,,
\\
{\rm m}_{ik}
	& := \frac{1}{\av{Y}}\int_{Y^1}\brcs{
			\delta_{ik}-\delta_{ij}\partial_{y_j}\xi^{3_k}(y)
		}\,dy\,,
\\
\epsilon^0_{ik}
	& := \frac{1}{\av{Y}}\int_{Y}\varepsilon(y)\brcs{
			\delta_{ik}-\delta_{ij}\partial_{y_j}\xi^{3_k}(y)
		}\,dy\,,
}{EfTe}
where
\bsplitl{
\ol{\rm v}_i
	:= \frac{1}{\av{Y^1}}\int_{Y^1}{\rm v}_i(y)\,dy
	\,.
}{vi}
Finally, the surface charge density per volume $\rho_s$ is determined by
\bsplitl{
\rho_s
	:= \frac{1}{\av{Y}}\int_{\partial Y^1\cap\partial Y^2}
		\sigma_{\rm s}(y)\,dy\,.
}{SuCh}
\end{thm}

\medskip

\begin{rem}\label{rem:VolAv}
i) Theorem \ref{thm:pmStPNP} can be immediately stated without fluid flow by setting ${\bf v}={\bf 0}$.
\\
ii) We note that it is not possible to derive (\ref{EffpmPNP}) by volume averaging 
or the representative volume method (RVM), since the system (\ref{PNP}) is nonlinear. Moreover, these approaches can not account for possible source terms or boundary conditions.
\end{rem}
\medskip

We prove this theorem by the multiscale expansion method with drift \cite{Marusic-Paloka2005,Allaire2010a}. 
The strength of this method is that it allows to systematically derive the physically relevant diffusion-dispersion 
relations (Taylor-Aris dispersion). We discuss this result and its physical implications in Section \ref{sec:Imp}.


\section{Formal derivation of Theorem \ref{thm:pmStPNP}}\label{sec:Pr}
Following along the lines of \cite{Schmuck2012}, we first define spatial differential 
operators arising by the multi-scale approach independent of fluid flow. These 
operators are
\bsplitl{
\begin{cases}
\quad{\cg A}_r\psi_r(x)
	:= -\sum_{i,j=1}^N\pd{}{x_i}\brkts{\varepsilon(x/r)\delta_{ij}\pd{\psi_r}{x_j}}
	&
\\\qquad\quad
	= \ebrkts{
		\brkts{
			s^{-2}{\cg A}_0
			+s^{-1}{\cg A}_1
			+{\cg A}_2
		}\psi
	}(x,x/r)\,,
	&
\\
\quad{\cg B}_r\psi_r(x)
	:= -\sum_{i,j=1}^N\pd{}{x_i}\brkts{\delta_{ij}\pd{\psi_r}{x_j}}
	&
\\\qquad\quad
	= \ebrkts{
		\brkts{
			s^{-2}{\cg B}_0
			+s^{-1}{\cg B}_1
			+{\cg B}_2
		}\psi
	}(x,x/r)
	\,,
	&
\\
\quad{\cg B}_r^\iota\psi_r(x)
	:= -\sum_{i,j=1}^N\pd{}{x_i}\brkts{z_\iota\psi_r\delta_{ij}\pd{{\rm u}^3_r}{x_j}}
	&
\\\qquad\quad
	= \ebrkts{
		\brkts{
			s^{-2}{\cg B}^\iota_0
			+s^{-1}{\cg B}^\iota_1
			+{\cg B}^\iota_2
		}\psi
	}(x,x/r)
	\qquad\textrm{for }\iota=1,2
	\,,
	&
\end{cases}
}{AsBs}
where ${\cg A}_i$ are related to the Poisson equation and for 
$i=0,1,2$ defined by,
\bsplitl{
{\cg A}_0
	& := -\sum_{i,j=1}^N\pd{}{y_i}\brkts{
		\varepsilon(y)\delta_{ij}\pd{}{y_j}
	}\,,
\\
{\cg A}_1
	& := -\sum_{i,j=1}^N\pd{}{x_i}\brkts{
		\varepsilon(y)\delta_{ij}\pd{}{y_j}
	}
	-\sum_{i,j=1}^N\pd{}{y_i}\brkts{
		\varepsilon(y)\delta_{ij}\pd{}{x_j}
	}\,,
\\
{\cg A}_2
	& :=	-\sum_{i,j=1}^N\pd{}{x_i}\brkts{
		\varepsilon(y)\delta_{ij}\pd{}{x_j}
	}\,,
}{A0A1A2}
The operators ${\cg B}^\iota_i$ and ${\cg B}^3_i$ related to the Nernst-Planck 
equations are defined for $r=1,2$ by,
\bsplitl{
{\cg B}_0
	& := -\sum_{i,j=1}^N\pd{}{y_i}\brkts{
		\delta_{ij}\pd{}{y_j}
	}\,,
\\
{\cg B}_1
	& := -\sum_{i,j=1}^N\pd{}{x_i}\brkts{
		\delta_{ij}\pd{}{y_j}
	}
	-\sum_{i,j=1}^N\pd{}{y_i}\brkts{
		\delta_{ij}\pd{}{x_j}
	}\,,
\\
{\cg B}_2
	& :=	-\sum_{i,j=1}^N\pd{}{x_i}\brkts{
		\delta_{ij}\pd{}{x_j}
	}\,,
\\
{\cg B}^\iota_0
	& := -z_\iota\sum_{i,j=1}^N\pd{}{y_i}\brkts{
		\delta_{ij}\pd{{\rm u}_0^3}{y_j}
	}\,,
\\
{\cg B}^\iota_1
	& := -z_\iota\sum_{i,j=1}^N\pd{}{x_i}\brkts{
		\delta_{ij}\pd{{\rm u}_0^3}{y_j}
	}
	-z_\iota\sum_{i,j=1}^N\pd{}{y_i}\brkts{
		\delta_{ij}\pd{{\rm u}_0^3}{x_j}
	}
	-z_\iota\sum_{i,j=1}^N\pd{}{y_i}\brkts{
		\delta_{ij}\pd{{\rm u}_1^3}{y_j}
	}
	\,,
\\
{\cg B}^\iota_2
	& :=	-\sum_{i,j=1}^N\pd{}{x_i}\brkts{
		\delta_{ij}\pd{{\rm u}_0^3}{x_j}
	}
	-z_\iota\sum_{i,j=1}^N\pd{}{x_i}\brkts{
		\delta_{ij}\pd{{\rm u}_1^3}{y_j}
	}
\\&\quad
	-z_\iota\sum_{i,j=1}^N\pd{}{y_i}\brkts{
		\delta_{ij}\pd{{\rm u}_1^3}{x_j}
	}
	-z_\iota\sum_{i,j=1}^N\pd{}{y_i}\brkts{
		\delta_{ij}\pd{{\rm u}_2^3}{y_j}
	}
	\,.
}{BrB3}
The definitions \reff{AsBs}, \reff{A0A1A2}, and \reff{BrB3} allow to 
obtain the following sequence of problems for periodic flow as described 
in \reff{MiPr} by equating terms of equal power in $r$ , i.e.,
\bsplitl{
\mathcal{O}(r^{-2}):\quad
\begin{cases}
{\cg B}_0{\rm u}_0^\iota + {\cg B}_0^\iota{\rm u}_0^\iota
	+{\rm Pe}_{loc}({\bf v}\cdot\nabla_y){\rm u}_0^\iota
	= 0
	&\textrm{in } Y^1\,,
\\\quad
\textrm{${\rm u}_0^\iota$ is $Y^1$-periodic}\,,&
\\
{\cg A}_0 {\rm u}_0^3
	= 0
	&\textrm{in }Y\,,
\\\quad
\textrm{${\rm u}_0^3$ is $Y$-periodic}\,,&
\end{cases}
}{O-2}
{\bsplitl{
\mathcal{O}(r^{-1}):\quad
\begin{cases}
\brkts{{\cg B}_0 + {\cg B}_0^\iota}{\rm u}_1^\iota
	+{\rm Pe}_{loc}({\bf v}\cdot\nabla_y){\rm u}_1^\iota
\\\qquad
	= -\brkts{{\cg B}_1 + {\cg B}_1^\iota}{\rm u}_0^\iota
	+({\bf v}^*-{\rm Pe}_{loc}{\bf v})\cdot\nabla_x{\rm u}_0^\iota
	&\textrm{in } Y^1\,,
\\\quad
\textrm{${\rm u}_1^\iota$ is $Y^1$-periodic}\,,&
\\
{\cg A}_0 {\rm u}_1^3
	= -{\cg A}_1 {\rm u}_0^3
	&\textrm{in }Y\,,
\\\quad
\textrm{${\rm u}_1^3$ is $Y$-periodic}\,,&
\end{cases}
}{O-1}
{\bsplitl{
\mathcal{O}(1):\quad
\begin{cases}
\brkts{{\cg B}_0 + {\cg B}_0^\iota}{\rm u}_2^\iota
	+{\rm Pe}_{loc}({\bf v}\cdot\nabla_y){\rm u}_2^\iota
\\\qquad
	= -\brkts{{\cg B}_1 + {\cg B}_1^\iota}{\rm u}_1^\iota
	+{\rm Pe}_{loc}(\ol{\bf v}-{\bf v})\cdot\nabla_x{\rm u}_1^\iota
\\\qquad
	-\brkts{{\cg B}_2 + {\cg B}_2^\iota}{\rm u}_0^\iota
	-\partial_t {\rm u}_0^\iota
	&\textrm{in } Y^1\,,
\\\quad
\textrm{${\rm u}_2^\iota$ is $Y^1$-periodic}\,,&
\\
{\cg A}_0 {\rm u}_2^3
	= \brkts{{\rm u}_0^1-{\rm u}_0^2}
	-{\cg A}_1 {\rm u}_1^3
	-{\cg A}_2 {\rm u}_0^3
	&\textrm{in }Y\,,
\\\quad
\textrm{${\rm u}_2^3$ is $Y$-periodic}\,.&
\end{cases}
}{O-0}
Using \reff{A0A1A2}, \reff{BrB3} and the interfacial boundary conditions, we can rewrite \reff{O-2} more 
intuitively by,
\bsplitl{
\begin{cases}
-\Delta_y{\rm u}_0^1
	-{\rm div}_y\brkts{z_1{\rm u}_0^1\nabla_y{\rm u}_0^3}
	+{\rm Pe}_{loc}({\bf v}\cdot\nabla_y){\rm u}_0^1
	= 0
	&\textrm{in }Y^1\,,
\\\quad
	\pmb{\nu}\cdot\brkts{
		\nabla_y{\rm u}_0^1
		+z_1{\rm u}_0^1\nabla_y{\rm u}_0^3
		+{\rm Pe}_{loc}{\bf v}\cdot {\rm u}_0^1
	}
	= 0
	&\textrm{on }\partial Y^1\,,
\\
-\Delta_y{\rm u}_0^2
	-{\rm div}_y\brkts{z_2{\rm u}_0^2\nabla_y{\rm u}_0^3}
	+{\rm Pe}_{loc}({\bf v}\cdot\nabla_y){\rm u}_0^2
	= 0
	&\textrm{in }Y^1\,,
\\\quad
	\pmb{\nu}\cdot\brkts{
		\nabla_y{\rm u}_0^2
		+z_2{\rm u}_0^2\nabla_y{\rm u}_0^3
		+{\rm Pe}_{loc}{\bf v}\cdot {\rm u}_0^2
	}
	= 0
	&\textrm{on }\partial Y^1\,,
\\
-{\rm div}_y\brkts{\varepsilon(y)\nabla_y{\rm u}_0^3}
	= 0
	&\textrm{in }Y\,.
\end{cases}
}{IO-2}
It is now standard to deduce from \reff{IO-2} that
${\rm u}_0^1(x,y,t)={\rm u}_0^1(x,t)$ and ${\rm u}_0^2(x,y,t)={\rm u}_0^2(x,t)$. Next, we search for 
the explicit form of \reff{O-1}, i.e.,
\bsplitl{
\begin{cases}
-\Delta_y {\rm u}_1^\iota
	-{\rm div}_y\brkts{
		z_\iota{\rm u}_0^\iota\nabla_y {\rm u}_1^3
	}
	+{\rm Pe}_{loc}({\bf v}\cdot\nabla_y){\rm u}_1^\iota
	&
\\\qquad
	= 	
	{\rm div}_y\nabla_x{\rm u}_0^\iota
	+{\rm div}_y\brkts{
		z_\iota{\rm u}_0^\iota\nabla_x{\rm u}_0^3
	}
	+({\bf v}^*-{\rm Pe}_{loc}{\bf v})\cdot\nabla_x{\rm u}_0^\iota
	&\textrm{in }Y^1\,,
\\\quad
	\pmb{\nu}\cdot\nabla_y\bigl(
		{\rm u}_1^\iota
		+z_\iota{\rm u}^\iota_0\nabla_y{\rm u}_1^3
		+{\rm Pe}_{loc}{\bf v}{\rm u}_1^\iota
\\\qquad
	+ \nabla_x{\rm u}_0^\iota
	+z_\iota{\rm u}_0^\iota\nabla_x{\rm u}_0^3
	\bigr)
	= 0
	&\textrm{on }\partial Y^1\,,
\\
-{\rm div}_y\brkts{
		\varepsilon(y)\nabla_y{\rm u}_1^3
	}
	= {\rm div}_x\brkts{
		\varepsilon(y)\nabla_y{\rm u}_0^3
	}
	+{\rm div}_y\brkts{
		\varepsilon(y)\nabla_x{\rm u}_0^3
	}
	&\textrm{in }Y\,,
\end{cases}
}{IO-1}
where $\iota=1,2$. For $\iota=1,2,3$, we make the usual ansatz
\bsplitl{
{\rm u}_1^\iota
	= \sum_{j=1}^N\frac{\partial {\rm u}_0^\iota}{\partial x_j}(x,t)\xi^{\iota_j}(y)\,,
}{u1r}
which allows to rewrite \reff{IO-1}$_1$ and \reff{IO-1}$_2$ as 
\bsplitl{
\begin{cases}
-\sum_{i,j=1}^N\pd{}{y_i}\brkts{
		\delta_{ij}\pd{}{y_j} 
			\xi^{\iota_j}(y) 
	}
	+{\rm Pe}_{loc}({\bf v}\cdot\nabla_y)\xi^{\iota_j}
\\\qquad
	=-\sum_{i,j=1}^N\pd{}{y_i}\brkts{
		\delta_{ij}\pd{}{y_j} 
			\xi^{3_j}(y) 
	}
	+({\rm v}^*_j-{\rm Pe}_{loc}{\rm v}_j)
	&\textrm{in }Y^1\,,
\\\quad
	\pmb{\nu}\cdot\brkts{
		\brkts{
			\nabla_y\xi^{\iota_j}(y)-{\bf e}_i
		}
		+\brkts{
			\nabla_y\xi^{3_j}(y)-{\bf e}_i
		}
		+{\rm Pe}_{loc}{\bf v}\xi^{\iota_j}
	}=0
	&
	\textrm{on }\partial Y^1\,,
\\\quad
	\xi^{\iota_j}(y)\textrm{ is $Y$-periodic in $y$ and $\int_{Y^1}\xi^{\iota_j}(y)\,dy=0$}\,,	
\\
-\sum_{i,j=1}^N\pd{}{y_i}\brkts{
		\varepsilon(y)\delta_{ij}\pd{}{y_j} 
			\xi^{3_j}(y) 
	}
	= 0
	&\textrm{in } Y\,,
\\\quad
	\xi^{3_j}(y)\textrm{ is $Y$-periodic in $y$ and $\int_Y\xi^{3_j}(y)\,dy=0$}\,,
\end{cases}
}{IIO-1}
where we used the Assumption (AI). Next, we guarantee the solvability of \reff{IIO-1} via the following {\bf Fredholm 
alternative}:\\
\emph{
Up to an additive constant, the boundary value problem
\bsplitl{
\begin{cases}
{\rm Pe}_{loc}({\bf v}\cdot\nabla_y)w
	- \Delta_yw
	= h(y)
	&\textrm{in }Y^1\,,
\\
\pmb{\nu}\cdot\brkts{
		(\nabla_yw-{\bf e}_j)
		+{\rm Pe}_{loc}{\bf v}w
	}
	= g(y)
	&\textrm{on }\partial Y^1\,,
\\
w \textrm{ is $Y^1$-periodic}\,,
	&
\end{cases}
}{faBVP}
has a unique solution $w\in H^1(Y^1)$, if and only if the 
following compatibility condition holds
\bsplitl{
\int_{Y^1}h(y)\,dy
	= \int_{\partial Y^1} g(y)\,do(y)
	\,.
}{faCC}
}
Via \reff{IIO-1} we recognize that 
\bsplitl{
h(y) 
	& := 
	-\Delta_y\xi^{3_j}
	+({\rm v}_j^*-{\rm Pe}_{loc}{\rm v}_j)\,,
\\ 
g(y)
	& := 
	-\pmb{\nu}\brkts{
		\nabla_y\xi^{3_j}
		-{\bf e}_j
	}\,,	
}{gh}
such that the compatibility condition \reff{faCC} becomes
\bsplitl{
{\rm v}_j^*
	:= \frac{\rm Pe_{loc}}{\av{Y^1}}\int_{Y^1}{\rm v}_j(y)\,dy
	\,.
}{ccO-1}
Let us turn now to the last reference cell problem \reff{O-0} which has the 
explicit form
\bsplitl{
\begin{cases}
-\Delta_y {\rm u}_2^\iota
	+{\rm Pe}_{loc}({\bf v}\cdot\nabla_y){\rm u}_2^\iota
	&
\\\qquad
	={\rm div}_y\nabla_x{\rm u}_1^\iota
	+{\rm div}_x\nabla_y{\rm u}_1^\iota
	+{\rm div}_y\brkts{
		z_\iota{\rm u}_1^\iota\nabla_x{\rm u}_0^3
	}
	+{\rm div}_y\brkts{
		z_\iota{\rm u}_1^\iota\nabla_x{\rm u}_1^3
	}
\\\qquad
	+{\rm Pe}_{loc}(\ol{\bf v}-{\bf v})\cdot\nabla_x{\rm u}_1^\iota
	+\Delta_x{\rm u}^\iota_0
	+{\rm div}_x\brkts{
		z_\iota{\rm u}_0^\iota\nabla_x{\rm u}_0^3
	}
	+{\rm div}_x\brkts{
		z_\iota{\rm u}_0^\iota\nabla_y{\rm u}_1^3
	}
	&
\\\qquad
	+{\rm div}_y\brkts{
		z_\iota{\rm u}_0^\iota\nabla_x{\rm u}_1^3
	}
	+{\rm div}_y\brkts{
		z_\iota{\rm u}_0^\iota\nabla_y{\rm u}_2^3
	}
	-\partial_t{\rm u}_0^\iota
	&
	\textrm{in }Y^1\,,
\\\quad
\pmb{\nu}\cdot\brkts{
		\nabla_y{\rm u}_2^\iota
		-{\rm Pe}_{loc}{\bf v}{\rm u}_2^\iota
	}
	= 0
	&\textrm{on }\partial Y^1\,,
\\\quad
{\rm u}_2^\iota\textrm{ is $Y$-periodic in $y$}\,,
\\
-\Delta_y{\rm u}_2^3
	= ({\rm u}_0^1 - {\rm u}_0^2)
	+{\rm div}_x\nabla_y{\rm u}_1^3
	+{\rm div}_y\nabla_x {\rm u}_1^3
	+\Delta_x{\rm u}_0^3
	&\textrm{in }Y\,,
\\\quad
{\rm u}_2^3\textrm{ is $Y$-periodic in $y$}\,.
\end{cases}
}{IO-0}
The Fredholm alternative implies the following compatibility 
condition on \reff{IO-0}$_1$, i.e.,
\bsplitl{
&\int_{Y^1}{\rm Pe}_{loc}(\ol{\bf v}-{\bf v})\cdot\nabla_x{\rm u}_1^\iota
	+{\rm div}_x\nabla_y{\rm u}_1^\iota
	+\Delta_x{\rm u}_0^\iota
\\
	&\qquad
	-\partial_t {\rm u}_0^\iota
	+{\rm div}_x\brkts{
		z_\iota{\rm u}_0^\iota\nabla_x{\rm u}_0^3
	}
	+{\rm div}_x\brkts{
		z_\iota{\rm u}_0^\iota\nabla_y{\rm u}_1^3
	}\,dy
	= 0\,.
}{ccIO-0}
We rewrite equation \reff{ccIO-0} by its components such that we determine the 
upscaled diffusion and mobility tensors. We obtain
\bsplitl{
\frac{\av{Y^1}}{\av{Y}}\partial_t {\rm u}_0^\iota
	& -\sum_{i,j,k=1}^N \brkts{\frac{1}{\av{Y}}\int_{Y^1}\brcs{
		\delta_{ik}-\delta_{ij}\frac{\partial \xi^{\iota_k}}{\partial y_j}
	}\,dy}
	\frac{\partial^2 {\rm u}_0^\iota}{\partial x_i\partial x_k}
\\
	& -\sum_{i,k=1}^N \brkts{\frac{1}{\av{Y}}\int_{Y^1}{\rm Pe}_{loc}(\ol{\rm v}_i-{\rm v}_i)\xi^{\iota_k}
	\,dy}
	\frac{\partial^2 {\rm u}_0^\iota}{\partial x_i\partial x_k}
\\
	& -\sum_{i,k,j=1}^N \frac{1}{\av{Y}}\int_{Y^1}\frac{\partial}{\partial x_i}\brkts{
		z_\iota{\rm u}_0^\iota\brcs{
			\delta_{ik}-\delta_{ij}\frac{\partial \xi^{3_k}}{\partial y_j}
		}
	}\,dy
	= 0
	\,.
}{cc1IO-0}
The structure of equation \reff{cc1IO-0} suggests that we define the following effective diffusion-dispersion 
tensor $\hat{\rm D}:=\brcs{{\rm d}_{ik}}_{1\leq i,k\leq N}$ and mobility tensor 
$\hat{\rm M}:=\brcs{{\rm m}_{ik}}_{1\leq i,k\leq N}$, i.e.,
\bsplitl{
{\rm d}_{ik}
	& := \frac{1}{\av{Y}}\int_{Y^1}\brkts{
		\delta_{ik}\brkts{1+{\rm Pe}_{loc}(\ol{\rm v}_i-{\rm v}_i)\xi^{\iota_k}}
		-\sum_{j=1}^N\delta_{ij}\frac{\partial\xi^{\iota_k}}{\partial y_j}
	}\,dy
	\,,
\\
{\rm m}_{ik}
	& := \frac{1}{\av{Y}}\int_{Y^1}\brkts{
			\delta_{ik}-\sum_{j=1}^N\delta_{ij}\frac{\partial \xi^{3_k}}{\partial y_j}
	}\,dy\,.
}{effTe}
The non-standard form of the effective diffusion coefficient requires to verify the 
positive definiteness of the tensor $\hat{\rm D}$. In fact, it is enough to show the non-negativity 
of
\bsplitl{
\int_{Y^1}{\rm Pe}_{loc}(\ol{\rm v}_j-{\rm v}_j)\delta_{jk}\xi^{\iota_k}\,dy
	& = \sum_{i,j=1}^N \int_{Y^1}\delta_{ij}\frac{\partial }{\partial y_j}\xi^{\iota_j}
		\frac{\partial\xi^{\iota_k}}{\partial y_i}\,dy
	+{\rm Pe}_{loc}\int_{Y^1}{\bf v}\cdot\nabla_y\xi^{\iota_j}\,dy
\\
	&
	-\sum_{i,j=1}^N\int_{Y^1}\delta_{ij}\frac{\partial }{\partial y_j}\xi^{3_j}\frac{\partial}{\partial y_i}\xi^{\iota_k}\,dy
\\
	& = \sum_{i,j=1}^N \int_{Y^1}\delta_{ij}\frac{\partial }{\partial y_j}\xi^{\iota_k}
		\frac{\partial\xi^{\iota_k}}{\partial y_i}\,dy
	= \N{\nabla_y\xi^{\iota_k}}{L^2(Y^1)}^2
	\geq 0\,.
}{PoDe}
Hence, if we define the porosity as $\theta:=\frac{\av{Y^1}}{\av{Y}}$ and apply the definitions 
\reff{effTe} in \reff{cc1IO-0}, then we end up with effective macroscopic Nernst-Planck 
equations
\bsplitl{
\theta\frac{\partial}{\partial t}{\rm u}^\iota_0
	- {\rm div}\brkts{
		\hat{\rm D}\nabla {\rm u}_0^\iota
	}
	-{\rm div}\brkts{
		z_\iota{\rm u}_0^\iota\hat{\rm M}\nabla {\rm u}_0^3
	}
	= 0\,.
}{effNP}
It leaves to upscale the contributions of the 
surface charge \reff{MiPr}$_9$. To this end, we apply the surface two-scale convergence result 
established in \cite{Allaire1996,Neuss-Radu1996}. First, we write 
problem (\ref{MiPr})$_7$--(\ref{MiPr})$_9$ in the distributional sense, i.e., we multiply the 
Poisson equation (\ref{MiPr})$_7$ with 
$\varphi\in C^\infty(\Omega)$ and after integration over $\Omega$, we end up with the formulation
\bsplitl{
-\brkts{\varepsilon(x/r)\nabla\phi_r,\nabla\varphi}_\Omega
	& = \brkts{{\rm c}^+_r-{\rm c}^-_r,\varphi}_\Omega
	-\int_{I^r}
		\hat{\varepsilon}(x/r)\nabla\phi_r 
		\varphi{\bf n}\,do(x)\,.
}{DiFo}
With the boundary condition (\ref{MiPr})$_9$, the equation (\ref{DiFo}) reduces 
to
\bsplitl{
-\brkts{\varepsilon(x/r)\nabla\phi_r,\nabla\varphi}_\Omega
	& = \brkts{{\rm c}^+_r-{\rm c}^-_r,\varphi}_\Omega
	+r\int_{I^r}\sigma_{\rm s}\varphi\,do(x)\,,
}{FiChaFo}
where the boundary conditions on $I^r$ need to be scaled by $r$ as motivated in \cite{Allaire1996,Neuss-Radu1996}. 
Next, we pass to the limit $r\to 0$ in the first two terms in \reff{FiChaFo} as in \cite{Schmuck2010}. In the remaining term, 
we apply the test function 
$\varphi_r(x) := \varphi(x) +r\varphi_1(x,\frac{x}{r})$ with $\varphi\in H^1_0(\Omega)$ and 
$\varphi_1(x,y)\in L^2(\Omega;H^1_\sharp({\rm div},Y^1))$. We conclude via \cite[Theorem 2.1]{Allaire1996} that
\bsplitl{
r\int_{I^r}\sigma_{\rm s}(x/r)\varphi_r\,do(x)
	\to \frac{1}{\av{Y}}\int_\Omega\varphi(x)
		\int_{\partial Y^1\cap\partial Y^2}\sigma_{\rm s}(y)\,dxdo(y)\,,
}{bndryIn}
in the limit $r\to 0$ where we used that $r\int_{I^r}\av{\sigma_{\rm s}(x/r)}^2\,do(x)\leq C$ holds for a uniform 
$C>0$ under the assumption $\sigma_{\rm s}(y)\in L^\infty_\sharp(\partial Y^1\cap\partial Y^2)$. This finally leads to the upscaled Poisson equation \reff{MiPr}$_9$.

\section{Physical aspects of the effective porous media system \reff{EffpmPNP}}\label{sec:Imp}
\subsection{Einstein's relation and the mean-field approximation}\label{sec:EiSt}
The upscaled PNP equations demonstrate that Einstein's relation between diffusion $D$ and mobility $M$
coefficient, i.e., $M=\frac{D}{kT}$,  which holds for the microscopic equations, does not hold with respect to the porous media correction tensors $\hat{\rm D}$ 
and $\hat{\rm M}$, except in the special case of an insulating solid matrix without fluid flow (discussed below). At first, this may appear to be physically inconsistent, since we seem to lose the gradient flow structure 
\reff{NP}, as well as the Boltzmann distribution for ion densities in thermodynamic equilibrium with the electrostatic potential. However, there are several ways to understand this mathematical result  and its validity in physical terms. 

One physical interpretation is that the tensors $\hat{\rm D}$ and $\hat{\rm M}$ 
are not corrections of the transport coefficients, as generally assumed and proclaimed in homogenization theory, but rather corrections of the gradient operators.  This view is consistent with the engineering notion of tortuosity (discussed below) as a rescaling of the physical length for transport, only we see that for the full PNP equations, there is a separate correction tensor to define each ``mean field gradient".   
To clarify this perspective, we may define the mean field approximations 
\bsplitl{
\ol{\nabla c^\pm}	
	& :=\hat{\rm D}\nabla c^\pm\,,
\\
\ol{\nabla\phi}
	& :=\hat{\rm M}\nabla\phi\,,
}{MfAp}
which lead in the dimensional case to
\bsplitl{
D_\pm\hat{\rm D}\nabla c^\pm+kTz_\pm M_\pm \hat{\rm M}\nabla\phi
	& = D_\pm\ol{\nabla c^\pm}+kTz_\pm M_\pm \ol{\nabla\phi} 
\\&
	= kT M_\pm\ol{\nabla\mu}_\pm\,.
}{Einst}
Moreover, via \reff{MfAp} we define the mean field gradient of the 
chemical potential by
\bsplitl{
\ol{\nabla\mu}_\pm
	& := 
	\frac{\delta\mu}{\delta c^\pm}\ol{\nabla c^\pm}	
	+ z_\pm\frac{\delta\mu}{\delta \phi}\ol{\nabla\phi}
	= \frac{1}{kT} \frac{\delta\mu}{\delta c^\pm}\hat{\rm D}\nabla c^\pm	
	+ \frac{\delta\mu}{\delta \phi}\hat{\rm M}\nabla\phi\,,
}{MfChPo}
where $\frac{\delta\mu}{\delta\phi}$ and $\frac{\delta\mu}{\delta c^\pm}$ denote the variational derivative of 
$\mu$. This allows us to recover the gradient flow \reff{NP} for dimensional quantities, i.e.,
\bsplitl{
\partial_tc^\pm
	= - {\rm div}\brkts{- c^\pm M\ol{\nabla\mu_\pm}}\,,
}{GrFl}
where $M=M_\pm =\frac{D_\pm}{kT}$. 
We remark that the mean field approximation \reff{MfChPo} only makes sense with respect to 
the gradient of the chemical potential. This fact is a direct consequence of the asymptotic 
two-scale expansion method. Formally, therefore, the chemical potentials $\mu_\pm$ remain 
unchanged, and hence Boltzmann's distribution for the ion densities still holds in thermodynamic equilibrium for the assumed dilute solution.
Formula \reff{MfChPo} also provides a general framework for obtaining mean field approximations of arbitrary diffusional chemical potentials that are defined variationally from free energy functionals for concentrated solutions~\cite{bazant2013,ferguson2012}.

It may still seem problematic that the upscaled equations do not always predict Boltzmann's distribution in steady state, but this finding also makes sense from the perspective of statistical averaging. If a nonlinear relationship such as Boltzmann's distribution, $c = \exp(-z e \phi/kT)$, holds at every microscopic point, then there is no guarantee that the same relationship holds for various reasonable definitions of spatially averaged variables, $\langle c \rangle \neq \exp( - z e \langle \phi \rangle / kT)$.   Insisting that this relationship holds (along with Einstein's relation) at the macroscopic scale is a particular mean-field approximation, which happens to differ from that provided by mathematical homogenization theory. The distinction lies in the way the statistically averaged concentration and potential variables and their gradients are defined.

Einstein's relation holds (by construction) whenever the electrostatic potential is defined as the ``potential of mean force". The total ionic flux can then be written as the gradient of an electrochemical potential, which has the following general definition in a concentrated solution,
\begin{equation}
\mu = \mu_{ref} + k T \ln a + z e \Phi.   
\label{eq:mudef}
\end{equation}
where $a$ is the chemical activity, measuring the free energy change from a reference state of chemical potential $\mu_{ref}$ when $\Phi=0$.  This form is the only one consistent with thermodynamics and leads to the Donnan equilibrium potential across a membrane between two electrolytes and the Nernst equation for the equilibrium voltage of a Faradaic charge-transfer reaction at an electrode~\cite{bazant2013}.   Equilibrium corresponds to $\mu=$constant with the generalized Boltzmann distribution, $a \propto \exp( -ze\Phi/kT)$.  In a uniform dilute solution, concentration replaces activity, and $\Phi=\phi$ is electrostatic potential of mean force. 

After homogenization in a dielectric porous medium,  the apparent breakdown of Einstein's relation implies the need to redefine either the chemical activity $a$ or the mean potential $\Phi$, so that Eq. (\ref{eq:mudef}) and the Einstein relation still hold.  In the general case where the diffusivity and mobility tensors are different, this cannot be accomplished simply by redefining the activity, because it is a scalar. However, it can be done by redefining the mean electric field,
\begin{equation}
\mathbf{E}^\prime_0 = - \nabla \phi^\prime_0 \equiv - \hat{D}^{-1} \hat{M} \nabla \phi_0 = \hat{D}^{-1} \hat{M} \mathbf{E}_0
\end{equation}
Here, $\Phi=\phi^\prime_0$ is a proper potential of mean force because the total ionic flux can now be written as $-\hat{D} c_0 \nabla \mu_0$, where  $\mu_0=kT \ln c_0 + ze \phi^\prime_0$, as in (\ref{eq:mudef}), and the Einstein and Boltzmann relations are satisfied. Inserting this transformation into the upscaled Poisson equation and its boundary conditions implies the following redefinition of the effective permittivity tensor, 
\begin{equation}
\hat{\varepsilon}^\prime = \hat{\varepsilon} \hat{M}^{-1} \hat{D}   \label{eq:epsdef}
\end{equation}
so that the upscaled Maxwell displacement field remains unchanged, 
\begin{equation}
\mathbf{D}_0 =  - \hat{\varepsilon} \nabla \phi_0 = - \hat{\varepsilon}^\prime \nabla \phi^\prime_0 = \mathbf{D}^\prime_0.
\end{equation}
The displacement vector is the fundamental quantity appearing in Maxwell's equations for the homogenized porous medium.

\subsection{Insulating porous matrix without flow}\label{sec:InPoMa}
The preceding discussion of the mean-field approximation becomes more clear in the special case of an insulating porous matrix without flow ($\bf v = 0$). In that case,  formally passing to the limit $\alpha\to 0$, the 
the porous media correction $\hat{\varepsilon}(\epsilon,\alpha)$ 
for the Poisson equation can be reduced to the classical diffusion (Laplace) corrector 
$\hat{\rm D}$, and the correction tensors for the mobility  $\hat{\rm M}:=\brcs{{\rm m}_{kl}}_{1\leq k,l\leq N}$ 
and $\hat{\rm D}:=\brcs{{\rm d}_{kl}}_{1\leq k,l\leq N}$ are also the same.  As a result, Einstein's relation holds for the upscaled equations, and Boltzmann's distribution is recovered in equilibrium.  This makes sense physically since the electric field and ionic fluxes are confined to the same tortuous pore space, where the assumed microscopic model of dilute solution theory upholds these relations. 

In contrast, if the porous matrix is a dielectric or conducting material, then the electric field spills into the matrix and leaves from the pores where the ions are confined. As such, the ions only sample part of the electrostatic potential in a given volume, and gradients of the upscaled electrostatic potential that averages over both the pores and the matrix do not properly capture the mean electrostatic forces on the ions. This is the simple physical reason that upscaling violates Einstein's relation and Boltzman's distribution in the general situation and requires redefinition of the potential and permittivity tensor in order to recover these basic relationships at the macroscopic scale.

\subsection{ Material tensor}
In \cite[equation (3.14)]{Schmuck2010}, it is shown for ${\bf v}={\bf 0}$ that the definitions in \reff{EfTe} represent a 
so-called effective ``material tensor": 
\bsplitl{
\hat{\rm S}({\bf u}):={\rm S}_{i_kj_l}({\bf u}) := 
\ebrkts{
\begin{array}{ccc}
{\rm d}_{kl} 
& 
0 
& {\rm u}^1{\rm m}_{kl}  
\\
0 
& {\rm d}_{kl}  
& -{\rm u}^2 {\rm m}_{kl}  
\\
0 
& 0 
& {\epsilon}^0_{kl}(\epsilon,\alpha)  
\end{array}
}\,,
}{HoTeEx}
for the field vector ${\bf u}:=[c^+,c^-,\phi]'$ and the 
right-hand side ${\bf I}({\bf u}):=[0,0,{\rm u}^1-{\rm u}^2]'$ by
\bsplitl{
\pmb{\partial}_{\bf t} {\bf u}
	- {\bf div}\brkts{\hat{\rm S}({\bf u})\pmb{\nabla}{\bf u}}
	= {\bf I}({\bf u})\,,
}{CoNo}
where $\pmb{\partial}_{\bf t}$ is the operator
\bsplitl{
\pmb{\partial}_{\bf t}
	:= \begin{bmatrix}
	\partial_t & 0 & 0 \\
	0 & \partial_t & 0 \\
	0 & 0 & 0 
	\end{bmatrix}\,,
}{TiDeOp}
and also $\pmb{\nabla}$ and $\pmb{\rm div}$ are correspondingly defined.   

In the case of an insulating matrix, the material tensor \reff{HoTeEx} simplifies to
\bsplitl{
\hat{\rm S}:={\rm S}_{i_kj_l}({\bf u}) := 
\ebrkts{
\begin{array}{ccc}
{\rm d}_{kl} 
& 
0 
& {\rm u}^1{\rm d}_{kl}  
\\
0 
& {\rm d}_{kl}  
& -{\rm u}^2{\rm d}_{kl}  
\\
0 
& 0 
& \epsilon^2{\rm d}_{kl}  
\end{array}
}\,.
}{HoTeEx2}
In this case, it becomes clear that upscaling is equivalent to rescaling of the coordinates, as in the engineering concept of tortuosity discussed below in Section \ref{sec:Tor}. 

Let us introduce the following coordinate tranformation,
\bsplitl{
\hat{\rm D}^{1/2}\tilde{x} := x\,,
}{CoTr}
where components of $\tilde{x}$ admitting ``$\infty$'' are subsequently to be treated as parameters. 
We remark that the transformation \reff{CoTr} accounts for a finite separation of scales and can be 
generalized to the case of a continuum of scales by the idea of metric-based upscaling 
introduced in \cite{Owhadi2007}. 
With \reff{CoTr} the gradient $\nabla_x$ and the divergence operator ${\rm div}_x$ change with respect to 
the new coordinates as follows
\bsplitl{
\nabla_x = \hat{\rm D}^{-1/2}\nabla_{\tilde{x}}\,,
	\quad\textrm{and}\quad
	{\rm div}_x  
	= (\nabla_x)'
	= {\rm div}_{\tilde{x}} \hat{\rm D}^{-1/2}\,,
}{TrOp}
where $\hat{\rm D}^{-1/2}$ denotes the matrix square root of $\hat{\rm D}^{-1}$. 
Via \reff{TrOp}, the tensor \reff{HoTeEx2} can be written in this new 
coordinates $\tilde{x}$ in the case of an insulating porous matrix, i.e., $\alpha=0$, by
\bsplitl{
\hat{\cal S}(\tilde{\bf u}):=\hat{\cal S}_{i_kj_l}(\tilde{\bf u}) := 
\ebrkts{
\begin{array}{ccc}
1
& 
0 
& \tilde{{\rm u}}^1 
\\
0 
& 1 
& -\tilde{{\rm u}}^2
\\
0 
& 0 
& \epsilon^2  
\end{array}
}\,,
}{HoTeEx3}
where $\tilde{{\rm u}}^\iota(t,\hat{\rm D}^{1/2}\tilde{x})={\rm u}^\iota(t,x)$ for $\iota=1,2,3$.
Hence, the material tensor \reff{HoTeEx2} takes the same form in the new coordinates 
$\tilde{x}$ as the classical PNP equations for homogeneous media in the case of an 
insulating porous matrix. Moreover, the porous media equation \reff{CoNo} reads in the new 
coordinates as 
\bsplitl{
\pmb{\partial}_{\bf t} \tilde{\bf u}
	-{\bf div}_{\tilde{x}}\brkts{\hat{\cal S}(\tilde{\bf u})\pmb{\nabla}_{\tilde{x}}\tilde{\bf u}}
	= {\bf I}(\tilde{\bf u})\,.
}{TrCoNo}

\subsection{Solutions to particular reference cell problems: Straight and perturbed channels}\label{sec:StCh}
The main purpose of this section is to demonstrate that under restrictive conditions one can apply available results 
from the literature on homogenization of diffusion equations in order to compute the correction tensors of the 
upscaled and more complex PNP system. To 
this end, we need to assume that the electric potential 
$\phi$ only exists in the electrolyte phase like the salt and charge concentrations and that the pores form straight channels. 
This is the case of an insulating 
porous matrix (i.e., $\alpha\to 0$) as studied in Section \ref{sec:InPoMa}. Hence, we know that the complex correction tensor $\hat{\varepsilon}(\epsilon,\alpha)$ simplifies to the corrector $\epsilon^2\hat{\rm D}$ where $\hat{\rm D}$ is defined by the classical 
reference cell problem of the diffusion equation.\\

We consider the reference cell depicted in Figure \ref{fig:Cors1} left (in 2D). The 
porous media correction with respect to the diffusion 
can be written in the two-dimensional case as follows
\bsplitl{
\hat{\rm D} &= \ebrkts{
\begin{array}{ccc}
{\rm d}_{11} & 0 \\
0 & {\rm d}_{22}
\end{array}
}\,.
}{EfDiStCh}
Obviously, in the case considered we have, as in \cite{Auriault1993}, ${\rm d}_{11}=\theta$ and 
${\rm d}_{22}=0$.  A strightforward extension of the straight channel to dimension 
three is depicted in Figure \ref{fig:Cors1} right.

\begin{figure}
	\centering
	\includegraphics[height=5cm,width=6cm]{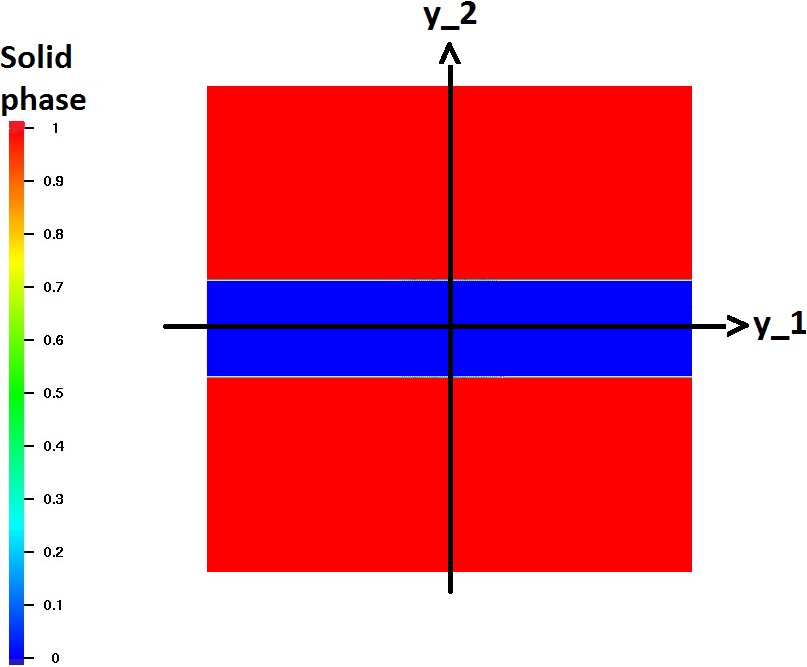}
	\hfill
	\includegraphics[height=5cm,width=6cm]{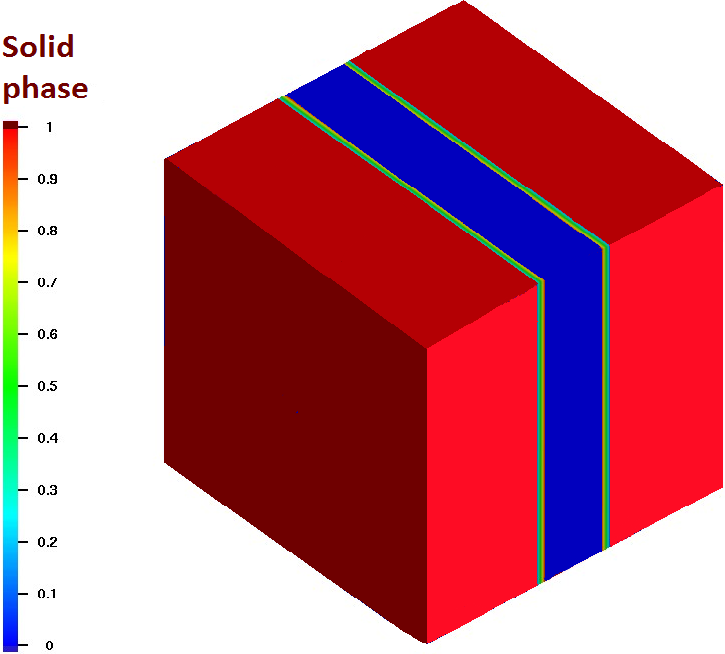}
	\caption{{\bf Example of straight channels:} \emph{Left:} Two-dimensional case. (pore phase is red) \emph{Right:} Three-dimensional case.}
	\label{fig:Cors1}
\end{figure}

\begin{figure}
	\centering
\includegraphics[height=6cm,width=6cm]{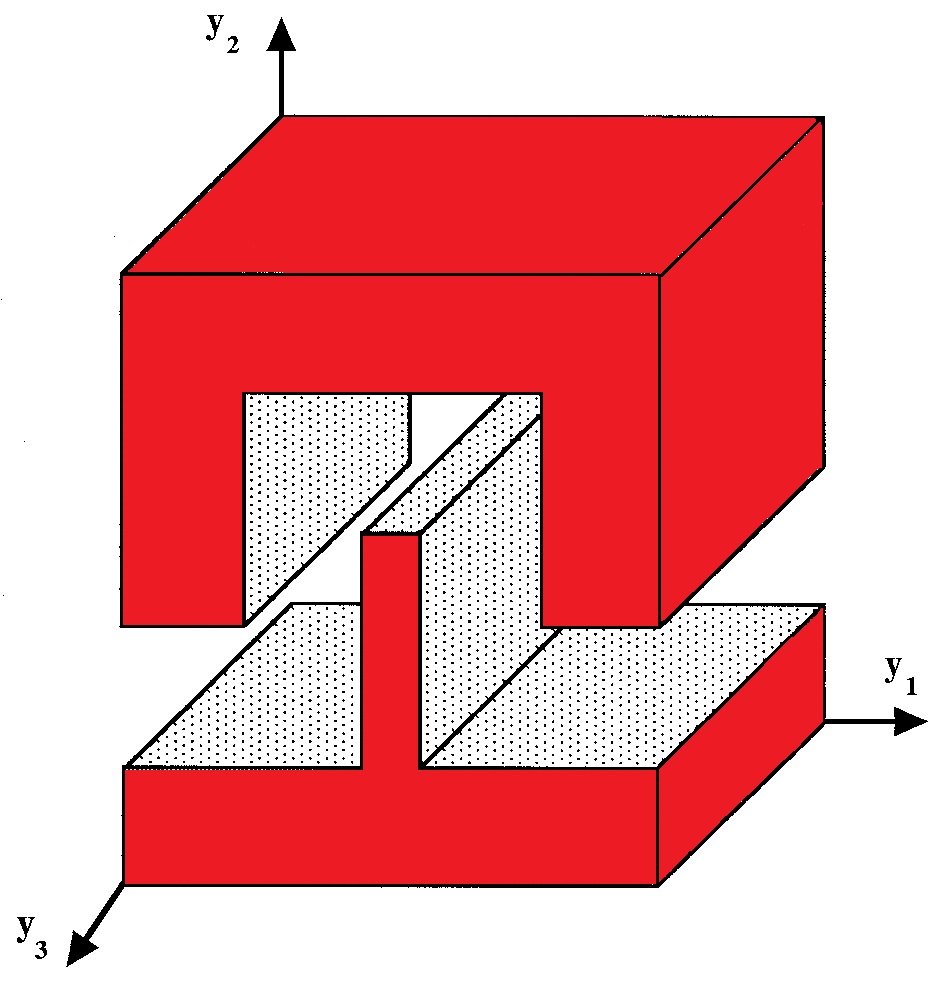}
	\hfill
\includegraphics[height=6cm,width=6cm]{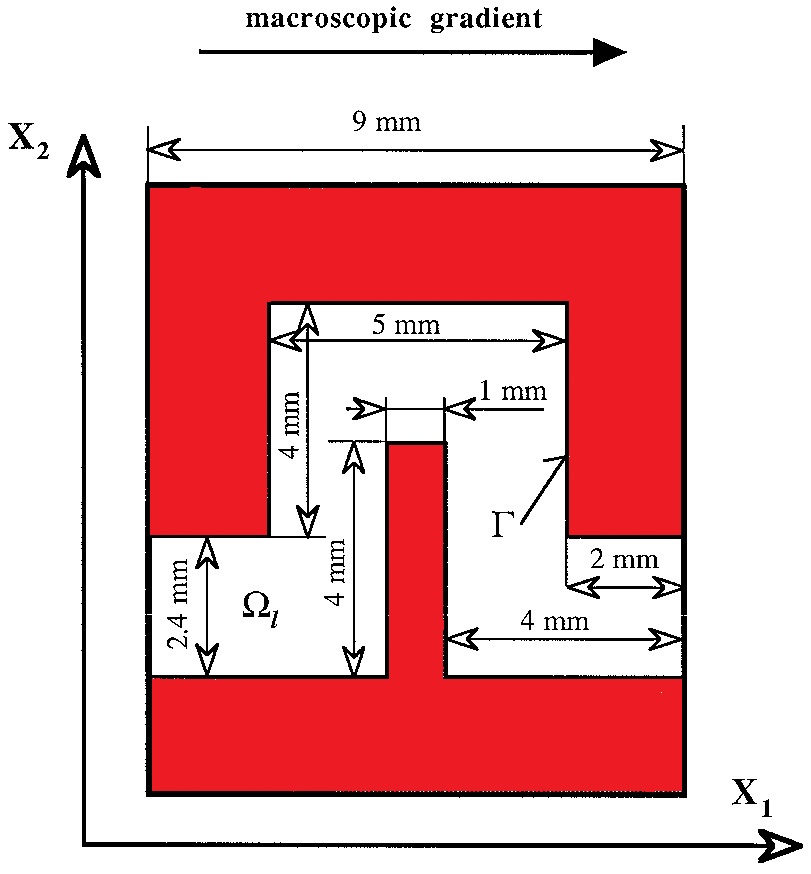}
	\caption{{\bf Perturbed straight channels in 3D, see \cite{Auriault1997}:} \emph{Left:} Reference cell geometry. \emph{Right:} Cross-section of the period.}
	\label{fig:PeCh3D}
\end{figure}
As opposed to straight channels, the case of perturbed straight channels 
requires the numerical calculation of the components $d_{ii}$ for $i=1,3$ of the effective 
diffusion tensor $\hat{\rm D}$ and for mathematical well-posedness, we have to set $\sigma_{\rm s}=0$, since 
$S=I^r\cap rY$ is not smooth in this case. The component ${\rm d}_{22}$ is $0$ as one would intuitively expect. We restate here briefly numerical results from \cite{Auriault1997} for the reader's convenience, i.e.,
\bsplitl{
\hat{\rm D}
	= \theta\ebrkts{
\begin{array}{ccc}
0.3833 & 0 & 0\\
0 & 0 & 0\\
0 & 0 & 1
\end{array}
}\,.
}{ptStCh}

\subsection{Tortuosity and effective diffusivity: A critical survey}\label{sec:Tor} In the following, 
we motivate that homogenization allows to validate current tortuosity relations and to give 
directions towards refinements of such relations. The explicit examples from Section \ref{sec:StCh} allow 
us to systematically understand the influence 
of the geometric structure on the tortuosity. Sometimes, the so-called diffusibility $Q$ is introduced to relate 
the molecular diffusion constant $D_{\rm f}$ and the effective diffusion constant $D_{\rm p}$ of a porous medium, i.e.,
\bsplitl{
D_{\rm p} = Q D_{\rm f}\,.
}{Q}
The expressions for $Q$ available in literature can be divided into three classes, see Brakel et al.~\cite{Brakel1974}: 
(1) \emph{Empirical correlations}, which express $Q$ as a function of the porosity $\theta$, 
i.e. $Q=f(\theta)$; (2) \emph{semi-empirical equations} based on a pore model where 
$Q$ is defined by the special class of functions $f(\theta)=\gamma\theta^\mu$ where the 
term $\theta^\mu$ is generally said to account for the influence of the smaller cross sectional 
surface available for diffusion; and (3) \emph{theoretical expressions} for $Q$ have been derived 
for dispersed solids in the form of spheres.

We first begin with a historical overview. In any porous system, the presence 
of solid particles/material causes the diffusion paths of species to deviate from straight lines. Consequently, 
the diffusion coefficients of species must be corrected. One tries 
to capture this deviation from straight lines in a porous medium by a term 
called tortuosity $\tau$, whose suitable definition is still an actual research 
topic.

By theory and dimensional reasoning, Petersen~\cite{Petersen1958} suggested that the 
diffusion coefficient is scaled by tortuosity as follows
\bsplitl{
D_{\rm p} = \frac{D_{\rm f}}{\tau^2}\,,
}{ClTo}
which implies $Q=1/\tau^2$. 
A similar relationship is introduced by Aris~\cite{Aris1969} and Satterfield~\cite{Satterfield1980}, i.e.
\bsplitl{
D_{\rm p} = \frac{\theta}{\tau}D_{\rm f}\,,
}{ArSaRel}
and hence $Q=\theta/\tau$. The simplest and most intuitive method to estimate $\tau$ (in the 2D-case of a single 
particle) is the ratio between the length of the real diffusion path $L_\gamma$ and the shortest distance of its end points $L_{ab}$, 
i.e., 
\bsplitl{
\tau :=\frac{L_\gamma}{L_{ab}}\,.
}{1DT}
In Brakel et al.~\cite{Brakel1974}, a slight generalization of (\ref{ClTo}) is considered by 
a constrictivity parameter ${\kappa}:=\brkts{\frac{D_{\rm p}}{\theta D_{\rm f}}}_{\tau=1}$, which 
accounts for the fact that the cross section of a segment varies over its length. Hence, (\ref{ClTo}) changes 
to
\bsplitl{
D_{\rm p} = \frac{\theta{\kappa}}{\tau^2}D_{\rm f}\,,
}{Constr}
so in this case, $Q=\frac{\theta \kappa}{\tau^2}$.

Further, Brakel et al.~\cite{Brakel1974} argued that for porous materials a function of the type $Q=f(\theta)$ 
does not exist. Moreover, they emphasize that the pragmatic value of the available $Q - \theta$ relations
is not very good. Recently, also Shen and Chen~\cite{Shen2007} gave a critical review 
of the current impact of tortuosity on diffusion. Therefore, we motivate our discussion and 
study of $Q$ in this section by suggesting a theoretically obtained $Q$ with the help of 
homogenization theory. The diffusibility $Q$ could turn out as a relevant parameter 
to compare empirical measurements with theoretically obtained effective quantities.

To this end, we first extend the above relations to tensorial versions, i.e., we denote 
by $\hat{\rm D}_{\rm p}$ the effective diffusion tensor in a porous environment and by 
$\hat{\rm D}_{\rm f}:=\brcs{D_f\delta_{ij}}_{ij}$ the 
molecular diffusion tensor in free space, where $\delta_{ij}$ denotes the Kronecker delta 
function. First, we extend (\ref{ClTo}) to
\bsplitl{
\hat{\rm D}_{{\rm p}}^{1/2}\hat{\tau} := 
		\hat{\rm D}_{{\rm f}}^{1/2}\,,
}{ToDiRe}
where $\hat{\rm D}_{{\rm p}}:=D_f\hat{\rm D}$ and the diffusion corrector $\hat{\rm D}$ is obtained 
by homogenization. We point out that the tensorial relation (\ref{ToDiRe}) also 
implies a tensorial diffusibility, i.e. $\hat{\rm Q} = 1/\hat{\tau}^2$. 

Another very interesting interpretation of \reff{ToDiRe} is possible in the case of an insulating 
porous matrix and normalized molecular diffusion $D_f=1$, see Section \ref{sec:InPoMa}. The tortuosity $\hat{\tau}$ in 
\reff{ToDiRe} corresponds then to the coordinate transformation \reff{CoTr}, i.e. $\tilde{x}=\hat{\tau}x$.

In view of (\ref{ArSaRel}) and (\ref{Constr}), we motivate further the extensions of (\ref{ArSaRel}) to 
\bsplitl{
\hat{\rm D}_{{\rm p}}^{1/2}\hat{\tau} 
	:= 
	\theta\hat{\rm D}_{{\rm f}}^{1/2}
		\,,
}{ToPoDiRe}
with corresponding $\hat{\rm Q}\hat{\tau} = \theta$ and the extension of \reff{ArSaRel}
\bsplitl{
\hat{\rm D}_{{\rm p}}^{1/2}\hat{\tau} 
	:= \brkts{
		\theta\hat{\kappa} \hat{\rm D}_{{\rm f}}
	        }^{1/2}\,,
}{ToPoStDiRe}
with $\hat{\rm Q}\hat{\tau}^2 = \theta\hat{\kappa}$, that is, we extended $\kappa$ towards a tensorial 
constructivity $\hat{\kappa}$.

\medskip

\emph{Comparison of the phenomenological relations \reff{ClTo}, \reff{ArSaRel}, and \reff{Constr} with the 
homogenized relations \reff{ToDiRe}-\reff{ToPoStDiRe}:}
Let us apply definition (\ref{ToDiRe}) to the examples from Sections 
\ref{sec:StCh}. In the case of straight channels, see 
Figure \ref{fig:Cors1} on the right-hand side, the definition (\ref{ToDiRe}) implies the following tortuosity tensor 
\bsplitl{
\hat{\tau}
	= \ebrkts{
\begin{array}{ccc}
1/\sqrt{\theta} & 0 & 0\\
0 & 0 & 0\\
0 & 0 & 1/\sqrt{\theta}
\end{array}
}\,.
}{StChTo}
We point out that the porosity $\theta$ with respect to straight channels corresponds to the channel height on 
the unit reference cell. Let us compare (\ref{StChTo}) with the intuitive definition (\ref{1DT}). If we apply definition (\ref{1DT}) in a straightforward manner, then $\tau=1$. However, it is not clear for straight channels, 
which path $L_\gamma$ is reasonable. Let us check for example the average 
\bsplitl{
L_\gamma=\frac{1}{n}\sum_{i=1}^nL_{\gamma_i}\,,
}{LgAv}
where $n\in\mathbb{N}$.
With (\ref{LgAv}), $n=3$, the path lengths $\gamma_1:=L_{ab}=1$, $\gamma_2:=L_{ab}+\theta=1+\theta$ where the porosity $\theta$ is the channel height, and $\gamma_3:=\sqrt{1+\theta^2}$  the diameter, we get $\tau=L_\gamma$ where $L_\gamma=\frac{1}{3}\brkts{\sqrt{1+\theta^2}+1+(1+\theta)}$. In following Boudreau~\cite[Section 2]{Boudreau1996}, the tortuosity must approach unity for $\theta\to 1$. This is violated by definition (\ref{1DT}) together with \reff{LgAv} since $\tau=1+\frac{\sqrt{2}}{3}$ with \reff{LgAv}. But the tortuosity \reff{StChTo} defined via the homogenization process perfectly satisfies this condition, see \cite[Section 2]{Boudreau1996}. 
Accordingly, in the case of perturbed straight channels as considered in Section \ref{sec:StCh}, the 
tortuosity tensor (\ref{ToDiRe}) becomes
\bsplitl{
\hat{\tau}
	= \ebrkts{
\begin{array}{ccc}
1/\sqrt{0.3833\theta} & 0 & 0\\
0 & 0 & 0\\
0 & 0 & 1/\sqrt{\theta}
\end{array}
}\,.
}{PeStChTo}
One immediately recognizes that $\hat{\tau}_{11}$ in \reff{PeStChTo} is $>1$ in the limit $\theta\to1$. Hence 
Boudreau~\cite[Section 2]{Boudreau1996} doesn't hold. These two contradictions advise caution when using definitions \reff{ClTo} 
and \reff{ToDiRe}.
Next, we examine the definition (\ref{ToPoDiRe}) which becomes for the case of straight channels
\bsplitl{
\hat{\tau}
	= \ebrkts{
\begin{array}{ccc}
1 & 0 & 0\\
0 & 0 & 0\\
0 & 0 & 1
\end{array}
}\,.
}{ToPoDi}
A comparison of (\ref{ToPoDi}) with (\ref{1DT}) shows perfect agreement, i.e. 
$\brkts{\tau_{(\ref{ToPoDi})}}_{11}=1=\tau_{(\ref{1DT})}$. However, in the case of perturbed straight 
channels, we depend on the numerical accuracy. Since the mesh in \cite{Auriault1997} is 
not very fine, we cannot necessarily expect equality. In fact, we obtain $\brkts{\tau_{(\ref{ToPoDi})}}_{11}=2.6$ 
and $\tau_{(\ref{1DT})}=\frac{4\cdot 4+1}{9}=1.9$. However, these discrepancies also motivate 
the critical statements of \cite{Brakel1974,Shen2007} about the pragmatic value of 
tortuosity as mentioned above. We leave the investigation of the definition (\ref{ToPoStDiRe}) to the 
interested reader, since the definition of the constrictivity parameter in \cite{Brakel1974} is a 
delicate point and again a new source for modeling errors.

As a conclusion of this discussion, we motivate that homogenization theory allows to derive effective 
equations which do not require a questionable tortuosity or diffusivity parameter. Moreover, these correction 
tensors provide a tool to check available tortuosity or diffusivity definitions and might suggest directions 
on how to improve their consistency. In view of 
Section \ref{sec:EiSt}, it seems also  relevant to recall that homogenization does not correct the diffusion constant 
but rather the spatial derivatives, i.e., the gradients.

\subsection{Ambipolar diffusion equation for a binary electrolyte}\label{sec:AmDi}
Motivated by the considerations in Mani and Bazant~\cite{Mani2010} by volume-averaging, we study here the equivalent problem for the homogenized equations \reff{EffpmPNP}, i.e., for ${\bf v}={\bf 0}$. The advantage of homogenization theory relies on the fact that we are able to accurately treat nonlinear terms. Up to now, there exists no general rule how to upscale the nonlinear terms by volume-averaging approaches. We already saw that the physics and the form of the upscaled system 
\reff{EffpmPNP} include explicit parameters/tensors which are not present in free space case. Despite this additional features, it is straightforward to derive the ambipolar diffusion equations, which account for arbitrary ionic valences, by starting with the effective macroscopic PNP system \reff{EffpmPNP}.\\

To this end, we extend the porous media approximation (\ref{EffpmPNP}) to a dilute, asymmetric binary electrolyte 
with arbitrary ionic charges, $q_\pm = \pm z_\pm e$ in this section. For simplicity, 
we assume constant diffusivities $D_\pm$ in the microstructure and denote the 
corresponding upscaled diffusivities and mobilities by $\mathbb{D}_\pm=D_\pm\hat{\rm D}$ and $\mathbb{M}_\pm=M_\pm \hat{\rm D}$. 
Without loss of generality, we consider a negative surface charge, i.e., $\rho_s<0$. Moreover, we work in the context 
of an insulating porous matrix ($\alpha=0$) and ${\bf v}={\bf 0}$ such that the porous media correction tensors 
satisfy $\hat{\rm M}=\hat{\rm D}$ and $\hat{\varepsilon}=\epsilon^2\hat{\rm D}$, see Section \ref{sec:InPoMa}.
We simplify now the Poisson-Nernst-Planck system by applying the usual conventions
\bsplitl{
0
	& = e\brkts{z_+c_+-z_-c_-} + \rho_s\,,
\\
c 
	& = \brkts{z_+c_+ + z_-c_-} +\frac{\rho_s}{e}\,,
}{AmCo}
where the first relation expresses quasi-neutrality for the case of surface charge. This assumption 
naturally arises here in view of the derived effective equations (\ref{EffpmPNP}) for fixed surface charge. 
However, in \cite{Manzanares1992,Ramirez2007,Szymczyk2010} such a neutrality condition has been suggested by pure 
physical reasoning.
Furthermore, we will not make use of the Nernst-Einstein equation (or simply Einstein relation) between the diffusivion tensors $\mathbb{D}_\pm=D_\pm\hat{\rm D}$ and mobility tensors 
$\mathbb{M}_\pm=M_\pm\hat{\rm D}$. Hence, the ambipolar diffusion equation 
derived under the above assumptions takes the form
\bsplitl{
\theta\partial_t c
	& = \ol{D}{\rm div}\brkts{\hat{\rm D}\nabla c}
	-\frac{\ol{z}}{e}{\rm div}\brkts{
	\rho_s\hat{\rm D}\nabla\tilde{\phi}
	}
	-\frac{D_+\ol{z}}{kTez_+M_+}{\rm div}\brkts{\hat{\rm D}\nabla\rho_s}
	\,,
}{pmAmDi}
where we used the relations
\bsplitl{
\ol{D} := \frac{z_+M_+D_-+z_-M_-D_+}{z_+M_++z_-M_-}
\qquad\textrm{and}\qquad
\ol{z} := \frac{2z_+z_-M_+M_-kT}{z_+D_-M_++z_-D_+M_-}\,.
}{CpmRels}
We remember that $\hat{\rm D}$ is defined by (\ref{EfTe}) for ${\bf v}={\bf 0}$, see Section \ref{sec:InPoMa}. \\

The correction tensors $\hat{\rm D}$ for straight channels and for 
perturbed straight channels (in Section \ref{sec:StCh}) allow to accordingly 
rewrite the ambipolar diffusion equation (\ref{pmAmDi}), which describes a 
porous material for a surface charge density $\sigma_s$. In view of the volume-averaged straight channels 
studied in \cite{Mani2010}, we only consider in the following the example from Section \ref{sec:StCh}. With (\ref{EfDiStCh}), the equation (\ref{pmAmDi}) immediately takes the form 
\bsplitl{
\partial_t c
	& = \ol{D}\partial^2_{x_1} c
	-\frac{\ol{z}}{e}\partial_{x_1}\brkts{
	\rho_s\partial_{x_1}\tilde{\phi}
	}
	-\frac{D_+\ol{z}}{kTez_+M_+}\partial^2_{x_1}\rho_s
	\,.
}{pmAmDiStCh}
Interestingly, the porosity parameter $\theta$ cancels out in (\ref{pmAmDiStCh}).

\subsection{ Thin double layers at macroscopic scale}\label{sec:TDL}
Recently, the thin-double-layer formulation for microchannels at the {\it microscopic} (channel or pore) scale has been formally extended to porous media by Mani and Bazant~\cite{Mani2010} by including the surface charge as a homogeneous background charge in the electroneutrality condition. The same approximation for thin double layers at the {\it macroscopic} scale can be found in classical membrane models~\cite{teorell1935,meyer1936,spiegler1971} for the limit of large background charge, which corresponds to strong counter-ion selectivity.  In the opposite limit of thin double layers at the pore scale, the  porous medium is weakly charged and behaves like a "leaky membrane"~\cite{yaroshchuk2012,dydek2013}, whose ion concentrations can be significantly depleted and enriched by the passage of current, since only a small fraction of the ions are involved in screening the surface charge.  

Without restricting the relative thickness of the double layers $\lambda_D$ relative to the pore scale $\ell$, we consider here the general limit of thin double layers compared to the macroscopic scale $L$ of the porous medium.  
A systematic analysis of this limit involves homogenization theory to accurately 
treat the nonlinear terms in (\ref{PNP}) and to account for fluid flow in \reff{MiPr}. A further advantage of the homogenization 
method is that the resulting system (\ref{EffpmPNP}) is not restricted to a special geometry and is rather valid for general porous structures defined by a periodic reference cell, e.g. Figure \ref{fig:MicMac}, which induces a tensors \reff{EfTe} defining 
mean field gradients as the main part of the upscaling. 
In the case of straight channels, an insulating porous matrix (i.e. $\alpha=0$), and ${\bf v}={\bf 0}$, the correction tensor $\hat{\rm D}$ can be analytically obtained, as in 
Section \ref{sec:StCh}, although  for more complicated geometries, such as irregular  
channels, the correction tensor $\hat{\rm D}$ must be calculated numerically.


In order to describe situations with thin electrical double layers compared to the macroscopic length of the porous medium, 
we consider the {\bf thin double layer limit} in (\ref{EffpmPNP}) rewritten for the salt $c:=\frac{c^++c^-}{2\overline{c}}$ and charge $\rho:=\frac{c^+-c^-}{2\overline{c}}$ variables. In the general case of a polarizable solid matrix, one immediately 
sees that the limit $\epsilon\to 0$, does not reduce the complexity of the macroscopic formulation, i.e., formally by setting  $\epsilon=0$ in 
$\hat{\varepsilon}(\epsilon,\alpha)$. However, if we pass to the joint limit 
$\epsilon,\alpha\to 0$, where the solid matrix is electrically insulating and ${\bf v}={\bf 0}$, then the porous media 
Poisson-Nernst-Planck system behaves like the 
classical PNP for $\epsilon\to 0$. That means, we obtain the following leading order bulk approximation for salt density $c$, charge density $\rho$, and electric potential $\phi$, i.e.,
\bsplitl{
0 = {\rm div}\brkts{c\hat{\rm D}\nabla\phi}\,,&
\\
\theta\partial_t c 
	= {\rm div}\brkts{\hat{\rm D}\nabla c}
	-{\rm div}\brkts{{\rho_s}\hat{\rm D}\nabla\phi}\,,&
\\
0 = \rho+\rho_s\,.&
}{LOBA}
The first equation expresses charge conservation in the quasi-neutral bulk solution by setting the divergence of the current to zero. The second equation expresses total salt conservation. This description of bulk electrolytes with thin double layers is very well known and forms the basis for classical theories of electrochemical transport~\cite{Newman2004}}, based on the assumption of quasi-electroneutrality in the electrolyte, $\rho=0$. The third equation, however, is different  and expresses quasi-electroneutrality of the entire porous composite, including not only the diffuse ionic charge $\rho$, but also the homogenized surface charge, $\rho_s$. 

Mani and Bazant~\cite{Mani2010} recently argued that the macroscopic electroneutrality condition, equation (\ref{LOBA}), generally holds in the limit of thin double layers at the pore scale. The physical reason is that the counter-ions screening the surface charge in a  thin double layer provide an extra surface conductivity, proportional to the total diffuse double-layer charge, which is acted on by the same tangential electric field as in the nearby bulk solution. If the double layers were not thin, the electric field would be strongly perturbed by the diffuse charge throughout the pore, and the extra counter ions could not be viewed as simply providing extra conductivity for bulk ion transport.  It would be interesting, but beyond the scope of this paper, to study this limit $\lambda \ll \ell$ systematically in the framework of homogenization theory.

\subsection{Optimizing conductivity in straight channels}\label{sec:optCond}
We are interested in finding the effective conductivity tensor $\hat{\sigma}(x)$ of a binary symmetric electrolyte inside of a porous 
domain $\Omega$ with corresponding surface $\partial\Omega$ and giving directions towards 
its optimization with respect to the pore geometry. 
In the following, we formally combine necessary physical equations and mathematical tools in order to obtain a 
conductivity tensor $\hat{\sigma}$ that depends on geometrical parameters. The ideas presented here should 
serve as a first motivation for deeper physical insights and for future research directions towards more 
rigorous Definitions and Theorems. 

We assume that the domain $\Omega$ is a porous medium with porosity $\theta$. For simplicity, 
we consider the pores to be straight (cylindrical) channels where the solid forms an insulating 
porous matrix with ${\bf v}={\bf 0}$, that means, $\hat{\rm D}=\hat{\rm M}=\hat{\varepsilon}$, see Section 
\ref{sec:InPoMa}. For a current density 
$\ol{J}$ together with the electrostatic equations ${\rm div}\,\ol{J} = 0$ and 
${\rm rot}\,\ol{E}=0$, where $\ol{E}=\nabla\phi$ and $\phi$ is a solution of \reff{EffpmPNP}$_2$, it holds that 
\bsplitl{
{\rm div}\brkts{\hat{\sigma}\nabla\phi}
	= 0\,,
}{CoRe}
where the constitutive relation $\ol{J} = \hat{\sigma}\ol{E}$ entered. Moreover, the upscaled Nernst-Planck equations \reff{EffpmPNP}$_1$ provide the current density $\ol{J}$ for a binary symmetric electrolyte, i.e.,
\bsplitl{
\ol{J} := \hat{\rm D}\nabla\rho + c\hat{\rm D}\nabla\phi\,.
}{CuDe}
Next, we determine the conductivity $\hat{\sigma}$ of the electrolyte with the current density from (\ref{CoRe}) and the Nernst-Planck flux \reff{CuDe}. Therefore, we replace $\rho$ in 
(\ref{CuDe}) by the Poisson equation (\ref{EffpmPNP})$_2$ with $\epsilon^2\hat{\rm D}$ 
instead of $\hat{\varepsilon}_{kl}(\epsilon,\alpha)$ as explained in Section \ref{sec:InPoMa}.
We obtain
\bsplitl{
\ol{J} 
	& = -\hat{\rm D}\nabla
   \brkts{
   {\rm div}
	\brkts{
		\epsilon^2\hat{\rm D}\nabla\phi
    	}
   +\rho_s 
   }
   + c\hat{\rm D}\nabla\phi\,,
}{CuDeRe}
The structure of equation (\ref{CuDeRe}) motivates to consider the eigenvalue problem for the Laplace operator, i.e.,
\bsplitl{
\begin{cases}
-\Delta_y u_i(y) + \lambda_i u_i(y) = 0 
	&\quad\textrm{in }Y^1\,,
\\
u_i(y) = 0
	&\quad\textrm{on }\partial Y^1\cap\partial Y^2\,.
\end{cases}
}{EvPr}
We remark that it is not immediately clear what kind of boundary conditions are required in (\ref{EvPr}). The boundary condition (\ref{EvPr})$_2$ has the advantage that it gives a lower bound \cite{Cheeger1970,Kawohl2003} on the first eigenvalue $\lambda_1$ in (\ref{EvPr}) for the geometry defined by the pore phase $Y^1$. We point out that instead of using the macroscopic Laplace operator ${\rm div}\brkts{\hat{\rm D}\nabla\phi}$, we apply the microscopic 
Laplace operator $\Delta_y=r^2{\rm div}_x\brkts{\hat{\rm D}\nabla}_x$ on the pore phase of the reference cell $Y^1$. This allows us to add information 
about the pore geometry to the problem.
Hence, the 
eigenvalue $\lambda_1$ depends on the pore geometry which is the striking point for our 
optimization goal. Since the self-adjoint eigenvalue problem \reff{EvPr} is a regular Sturm-Liouville problem, 
we can use its solutions $\brcs{u_i}_i$ to generate an orthonormal basis in $L^2(\Omega)$. Thus, for any 
function $f\in L^2(\Omega)$ we have
\bsplitl{
f = \sum_i^\infty\abrkts{f,u_i}u_i\,,
}{OrEx}
where equality is in the sense of $L^2$.

Now, we can choose $\hat{\rm D}$ as in Section \ref{sec:StCh} for 
straight channels, if we additionally assume that the electrostatic potential only exists in the 
electrolyte phase.  Hence, after choosing $f=\partial_{x_1}\phi$, the relation (\ref{CuDeRe}) becomes
\bsplitl{
J^1 = \theta\brkts{\sum_i^\infty\brkts{
    \frac{\epsilon^2}{r^2}\Delta_y+c
    }\abrkts{\partial_{x_1}\phi,u_i}u_i    
    - \partial_{x_1}\rho_s
    }\,,
}{CuDeRe2}
where equality holds again in the $L^2$-sense.
We can now approximate \reff{CuDeRe2} by only considering the first eigenvalue $\lambda_1$ of 
\reff{EvPr}. That means, we obtain
\bsplitl{
J^1 \approx \theta\brkts{\brkts{
    \frac{\epsilon^2}{r^2}\lambda_1+c
    }\abrkts{\partial_{x_1}\phi,u_1}u_1    
    - \partial_{x_1}\rho_s
    }\,.
}{CuDeRe3}
Since $\rho_s$ is independent of $x_1$, we get the following approximations for the conductivity, i.e.,
\bsplitl{
\sigma_{11}
	:= \sigma_{11}(\theta,\lambda_1,\epsilon,r,c)
	\approx \theta\brkts{\frac{\epsilon^2}{r^2}\lambda_1+c}\,.
}{CoEl}
The dimensionless Debye length $\epsilon$ in equation \reff{CoEl} indicates that surface conduction plays a central role 
in ion transport through porous structures. Hence, materials with higher heterogeneities improve 
the ionic conductivity in view of this equation.
\\
This means that the optimization of the conductivity in direction of the straight pores 
is achieved by increasing $\epsilon$ and $\lambda_1$ for given $\theta$, $c$ and $r$. With the help of Cheeger's 
number $h(\Omega^r)$, we have an additional tool for optimizing the conductivity with respect to geometry. Due to 
Cheeger~\cite{Cheeger1970} 
and Kawohl and Fridman~\cite{Kawohl2003}, it holds that
\bsplitl{
\lambda_1\geq \brkts{\frac{h(\Omega^r)}{2}}^2\,.
}{Cheeg}
\emph{Example 1: (Square)} For a square $S_a:=[-a,a]^2$, Cheeger's number can be determined explicitely by
$h(S_a)=\frac{4-\pi}{(4-2\sqrt{\pi})a}$. Moreover, we know that the first eigenvalue 
is $\lambda_1(S_1)=2\pi^2$. This indicates that the lower bound given by estimate 
(\ref{Cheeg}) is not too sharp. However, it allows at least to obtain first insights for 
possible directions towards optimization of the conductivity (\ref{CoEl}).\\
\emph{Example 2: (Rectangle)} For a rectangle $R_{a,b}:=[-a,a]\times[-b,b]$, one immediately gets the following Cheeger 
constant, see \cite{Kawohl2006}, 
\bsplitl{
h(R_{a,b})=\frac{4-\pi}{a+b-\sqrt{(a-b)^2+\pi ab}}\,.
}{RabChe}
Hence, in order to optimize the conductivity (\ref{CoEl}) for a rectangle shaped pore $R_{a,b}$, we have to maximize $h(R_{a,b})$ what is equivalent to the minimization 
of $a$ and $b$. If we assume that we are given a porous material of characteristic length $b=l$, then it immediately follows that $h$ is maximal after minimizing the channel hight $a>0$.

\section{Conclusion}\label{sec:Disc}

We have applied a systematic, formal homogenization procedure for thePoisson-Nernst-Planck equations (\ref{MiPr}) for ion transport in charge porous media. The resulting upscaled macroscopic equations (\ref{EffpmPNP}) have a similar form as the microscopic equations, except for three fundamental modifications: (i) The ionic diffusivities and mobilities, as well as the effective medium permittivity, become tensorial coefficients, which are explicitly connected to the microstructure by solving the periodic reference cell problem, (ii) the total surface charge per volume appears as an extra ``background  charge" in the upscaled Poisson equation, and (iii) the diffusion corrector accounts for so-called diffusion-dispersion 
relations induced by a dominant periodic fluid flow. The porous-medium PNP equations may find many applications in electrochemical and biological systems involving ion transport in charged porous media, where effects of fluid flow can be neglected. Simplified equations for the limits of thin or thick double layers may also be appropriate in many cases.

There are many interesting avenues for future work, building on these results. There is a substantial literature on rigorous bounds and approximations for the effective diffusivity or conductivity of  a composite medium~\cite{torquato2002}, related to solutions of Laplace's equation with flux matching interfacial conditions. It would be challenging and useful to derive analogous mathematical bounds and approximations for the effective diffusivities and mobilities of ions in a charged composite medium, which appear as tensorial coefficients in our porous-medium PNP equations. One might expect analogs of the Wiener bounds for anisotropic composites to hold for striped microstructures and analogs of the  Hashin-Shtrikman bounds for isotropic microstructures to hold for space-filling random sphere packings, although the appearance of an internal length scale for electrostatic interactions (the Debye screening length) complicates such simple geometrical constructions.

It would also be valuable to find simple ways to approximate the solution to the reference-cell problem and thus derive simplified expressions for the tensorial diffusivities and mobilities. In the limit of thin double layers, this could be done using surface conservations laws, which are effective boundary conditions on the neutral solution obtained by singular perturbation methods~\cite{Chu2007,biesheuvel2010}. In the opposite limit of thick double layers, regular perturbation methods might be applied to capture effects of diffuse charge variations in the microstructure.

We close by emphasizing the open challenge of deriving effective ion transport equations in more general situations using homogenization theory.  We have already commented on the extension to concentrated solution theories based on the local density approximation (for chemical interactions) and the mean-field approximation (for electrostatics)~\cite{Bazant2009}. Going beyond these approximations in the microscopic equations can lead to non-local Nernst-Planck integral equations~\cite{gillespie2002,gillespie2011} or higher-order Poisson equations~\cite{bazant2011}, whose upscaled form remains to be determined. Perhaps even more challenging, and more important for many applications, would be to predict the effects of general, non-periodic fluid flow on the homogenized PNP equations, coupled to the Navier-Stokes equations with electrostatic body forces. When large currents exceeding diffusion limitation are applied to charged porous media, it has been predicted theoretically~\cite{dydek2011} and confirmed experimentally~\cite{deng2013} that complex nonlinear electrokinetic phenomena arise, which cannot be described by Taylor-Aris dispersion~\cite{yaroshchuk2011}, or our homogenization approximation, due to the formation of non-equilibrium   ``fingers" of high and low salt concentration at the pore scale~\cite{rubinstein2013}.



\medskip


\end{document}